# Cheap Artificial AB-Mountains, Extraction of Water and Energy from Atmosphere and Change of Regional Climate

## Alexander Bolonkin
C&R, 1310 Avenue R, #F-6, Brooklyn, NY 11229, USA
T/F 718-339-4563, aBolonkin@juno.com, or aBolonkin@gmail.com. http://Bolonkin.narod.ru

## Abstract

Author suggests and researches a new revolutionary method for changing the climates of entire countries or portions thereof, obtaining huge amounts of cheap water and energy from the atmosphere. In this paper is presented the idea of cheap artificial inflatable mountains, which may cardinally change the climate of a large region or country. Additional benefits: The potential of tapping large amounts of fresh water and energy. The mountains are inflatable semi-cylindrical constructions from thin film (gas bags) having heights of up to 3 - 5 km. They are located perpendicular to the main wind direction.  Encountering these artificial mountains, humid air (wind) rises to crest altitude, is cooled and produces rain (or rain clouds). Many natural mountains are sources of rivers, and other forms of water and power production - and artificial mountains may provide these services for entire nations in the future. The film of these gasbags is supported at altitude by small additional atmospheric overpressure and may be connected to the ground by thin cables. The author has shown (in previous works about the AB-Dome) that this closed AB-Dome allows full control of the weather inside the Dome (the day is always fine, the rain is only at night, no strong winds) and influence to given region. This is a realistic and cheap method of economical irrigation, getting energy and virtual weather control on Earth at the current time.

**Key words:** Local weather control, gigantic film AB-Dome, converting a dry region to subtropics, converting desolate wilderness to a prosperous region, macroprojects. Induced rainfall, rainmaking, green power.

*Note:* Some background material in this article is gathered from Wikipedia under the Creative Commons license. Version 1 is submitted on 31 January 2008, the small corrected version 2 is submitted on 10 May 2008.

## Introduction

**1. Short History.** A particularly ambitious proposal evaluated in the Farmhand *Fencepost (Australia)* was to build a new mountain range (from soil and stone!) in the West of Australia, so as to create rain in the dry interior of Australia (Engineering the Weather, Part 10, p.115). The idea, first suggested by L.H Hogan in his book *Man Made Mountain*, is to build a very large mountain range 4 km tall, 10km wide at the base, with a 2km plateau at the top and covering a distance of 2,000km. Purpose is to change the dry climate in the West of Australia. A trench approximately 100 meters deep, 200km wide and 2,000km long would be necessary to provide the amount of material to build the mountain range ($400,000km^3$). It has been estimated that the construction of such a mountain would be more than $100,000 billion (HWA, 2003). Given Australia national budget is $178 billion, the price tag would be 500 times the national budget. If the new mountain were to create an extra 50 million ML of runoff each year, then the cost of the water would be more than $100,000 per ML ($100/kL) (HWA,2003). That was a visionary but impractical project proposal.

**Joseph Friedlander** (of Shave Shomron, Israel) suggested (2008) to the author to apply the AB-Dome to this problem. In this work, the author develops a theory for estimation and computation of the effect and cost of artificial inflatable AB-Mountains. The result is this article. This theory and computations may be applied to any particular project in any country.

**Common information**.

**2. Climate**. Climate is the average and variations of weather in a region over long periods of time. Climate zones can be defined using parameters such as temperature and rainfall.

**3. Precipitation.** *General information about precipitation*. The extant amount of water in Earth's hydrosphere in the current era is constant. The average annual layer of Earth's precipitation

is about 1000 mm or 511,000 km³. 21% of this (108,000 km³) falls on land and 79% (403,000 km³) on oceans. Most of it falls between latitudes 20° North and 20° South. Both polar zones collect only 4% of Earth's precipitation. The evaporation from the World-Ocean equals 1250 mm (450,000 km³). 1120 mm returns back as precipitation and 130 mm by river inflow. The evaporation from land equals 410 mm (61,000 km³), the precipitation is 720 mm. The land loses 310 mm as river flow to the oceans (47,000 km³). These are average data. In some regions the precipitation is very different (fig.1).

  In meteorology, **precipitation** (also known as one class of **hydrometeors**, which are atmospheric water phenomena) is any product of the condensation of atmospheric water vapor that is deposited on the earth's surface. It occurs when the atmosphere (being a large gaseous solution) becomes saturated with water vapour and the water condenses and falls out of solution (i.e., precipitates). Air becomes saturated via two processes, cooling and adding moisture.

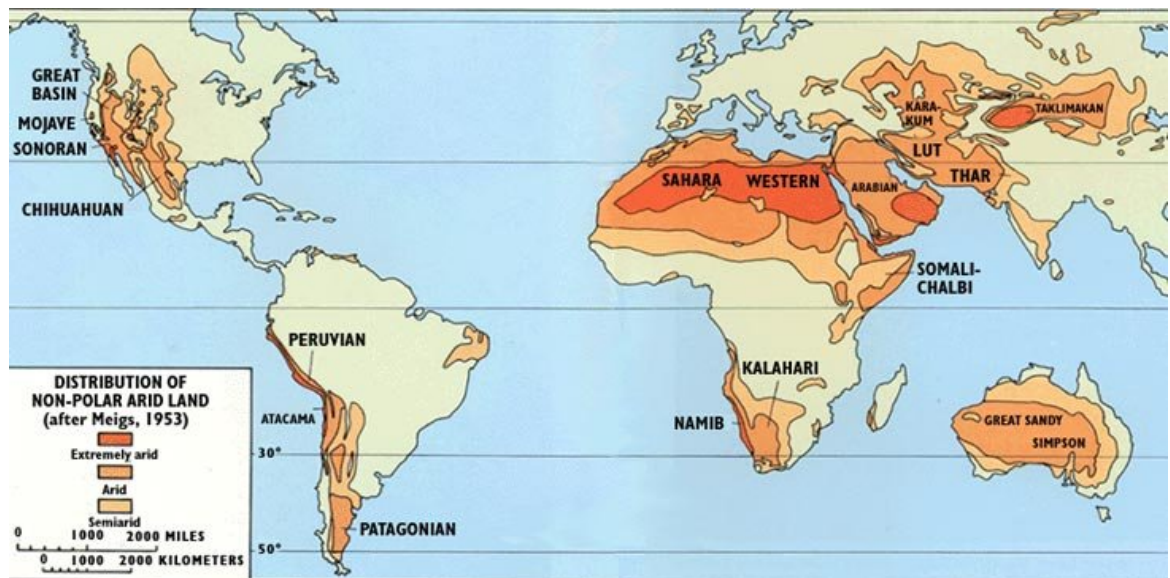

**Fig. 1**. Distribution non-polar arid land. 78 kb

   Precipitation that aches the surface of the earth can occur in many different forms, including rain, freezing rain, drizzle, snow, sleet, and hail. Virga is precipitation that begins falling to the earth but evaporates before reaching the surface. Precipitation is a major component of the hydrologic cycle, and is responsible for depositing most of the fresh water on the planet. Given the Earth's surface area, that means the globally-averaged annual precipitation is about 1 m, and the average annual precipitation over oceans is about 1.1 m.

   Air contains moisture, measured in grams of water per kilogram (or  m³) of dry air (g/kg) or g/m³, but most commonly reported as a relative humidity percentage. How much moisture a parcel of air can hold before it becomes saturated (100% relative humidity) depends on its temperature. Warmer air has a higher capacity for holding moisture than cooler air. Because of this property of the air, one way to saturate a parcel of air is to cool it. The dew point is the temperature that a parcel needs to be cooled to for saturation to occur.

Some cooling mechanisms include:

- Lift (convective, mechanical, positive vorticity advection)
    - Conductive cooling (warm air moves over a cool surface)
    - Radiational cooling (heat radiates off into space at night)
    - Evaporative cooling (air temperature falls as liquid water uses the energy to change phase to vapour).

The other way to saturate an air parcel is to add moisture to it, by:

- Precipitation falling from above (stratus forming in the rain under a higher cloud)
- Daytime heating evaporating water from the surface of oceans/lakes
  - Drier air moving over open water (snow streamers off the Great Lakes in winter)

**Fig.2.** Nature water cycle. 69 kb. Figure is deleted because article size is more 1 Mb.

**4. Orographic effects**. *Orographic precipitation* occurs on the windward side of mountains and is caused by the rising air motion of a large-scale flow of moist air across the mountain ridge, resulting in adiabatic cooling and condensation.

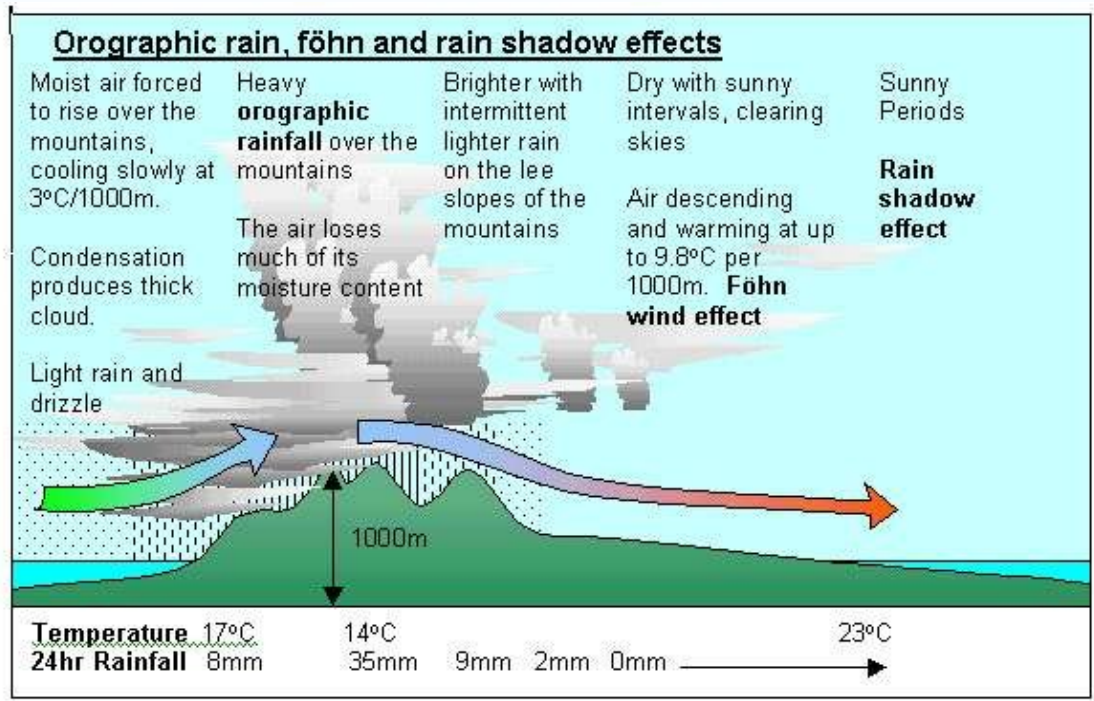

**Fig.3.** Orographic precipitation. After encounter with mountain the atmospheric temperature is 5 – 7 C more than the air temperature before encounter because the air temperature increases (when air pressure increases) at low altitude and vapor condences at high altitude. That air warming is an important benefit for countries which use the artificial mountains for protection from cold polar winds. 64 kb.

In mountainous parts of the world subjected to relatively consistent winds (for example, the trade winds), a more moist climate usually prevails on the windward side of a mountain than on the leeward (downwind) side. Moisture is removed by orographic lift, leaving drier air (see katabatic wind) on the descending (generally warming), leeward side where a rain shadow is observed.

Orographic precipitation is well known on oceanic islands, such as the Hawaiian Islands, where much of the rainfall received on an island is on the windward side, and the leeward side tends to be quite dry, almost desert-like, by comparison. This phenomenon results in substantial local gradients of average rainfall, with coastal areas receiving on the order of 500 to 750 mm per year (20 to 30 inches), and interior uplands receiving over 2.5 m per year (100 inches). Leeward coastal areas are especially dry 500 mm per year (20 inches) at Waikiki, and the tops of moderately high uplands are especially wet – ~12 m per year (~475 inches) at Wai'ale'ale on Kaua'i.

Highest Average Annual Precipication

| Contiment and Place | High everage precipitation, m/year (inch/year) | Elevation m (feet) |
|---|---|---|
| Asia, Mawsynram, India | 11.9 (467.4) | 1402 (4597) |
| Oceania, Mount Waialeale, Kaual | 11.7 (460) | 1570 (5138) |

| Australia, Relenden Ker, Qeensland | 8.64 (340) | 1556 (1556) |
|---|---|---|
| Europe, Crkvica, Rosnia - Herzegovina | 4.65 (184) | 1018 (1018) |

Twenty years ago, a record 683 inches of rain fell over a 12-month period on Mount Waialeale. It is arguably the wettest place on Earth, with an average annual rainfall of 460 inches. Here's why raindrops keep falling on this 5,208-foot peak.

Mount Waialeale received an average annual rainfall of 460 inches over a 32-year period. However, measurements over 38 years at Mawsynram, India, give it an edge of 7.4 inches per year; in 1860, nearby Cherrapunji recorded an unofficial 1,000 inches.
When warm air reaches a mountain range, it is lifted up the mountain slope, cooling as it rises.

This process is known as orographic lifting, a process common to all mountain ranges.
It is the main rainfall-producing mechanism in Hawaii. Here, orographic lifting of moisture-laden northeast tradewinds over the windward slopes of each island creates wet windward weather and dry leeward weather. Just 15 miles west of Mount Waialeale, the Kekaha coastal region receives less than 20 inches of rain a year. This condition is called a "rainshadow."

In South America, the Andes mountain range blocks most of the Atlantic moisture that arrives in that continent, resulting in a desert-like climate on the Pacific coast of Peru and northern Chile, since the cold Humboldt Current ensures that the air off the Pacific is dry as well. On the leeward side of the Andes is the Atacama Desert of Chile. It is also blocked from moisture by mountains to its west as well. Not coincidentially, it is the driest place on earth. The Sierra Nevada range creates the same effect in North America forming the Great Basin desert, Mojave Desert and Sonoran Desert.

5. **Wind**. Wind is the flow of air. More generally, it is the flow of the gases which compose an atmosphere; since wind is not unique to Earth. Simply it occurs as air is heated by the sun and thus rises. Cool air then rushes in to occupy the area the now hot air has moved from. It could be loosely classed as a convection current. Winds are commonly classified by their spatial scale, their speed, the types of forces that cause them, the geographic regions in which they occur, or their effect.

There are global winds, such as the wind belts which exist between the atmospheric circulation cells. There are upper-level winds which typically include narrow belts of concentrated flow called jet streams. There are synoptic-scale winds that result from pressure differences in surface air masses in the middle latitudes, and there are winds that come about as a consequence of geographic features, such as the sea breezes on coastlines or canyon breezes near mountains. Mesoscale winds are those which act on a local scale, such as gust fronts. At the smallest scale are the microscale winds, which blow on a scale of only tens to hundreds of meters and are essentially unpredictable, such as dust devils and microbursts.

Forces which drive wind or affect it are the pressure gradient force, the Coriolis force, buoyancy forces, and friction forces. When a difference in pressure exists between two adjacent air masses, the air tends to flow from the region of high pressure to the region of low pressure. On a rotating planet, flows will be acted upon by the Coriolis force, in regions sufficiently far from the equator and sufficiently high above the surface.

The three major driving factors of large scale global winds are the differential heating between the equator and the poles (difference in absorption of solar energy between these climate zones), and the rotation of the planet. Winds can shape landforms, via a variety of aeolian processes. Some local winds blow only under certain circumstances, i.e. they require a certain temperature distribution. Differential heating is the motive force behind land breezes and sea breezes (or, in the case of larger lakes, lake breezes), also known as on- or off-shore winds. Land absorbs and radiates heat faster than water, but water releases heat over a longer period of time. The result is that, in locations where sea and land meet, heat absorbed over the day will be radiated more quickly by the land at night, cooling the air. Over the sea, heat is still being released into the air at night, which rises. This convective motion draws the cool land air in to replace the rising air, resulting in a land breeze in the late night and early morning. During the day, the roles are reversed. Warm air over the land rises, pulling cool air in from the sea to replace it, giving a sea breeze during the afternoon and evening.

**Mountain breezes** and **valley breezes** are due to a combination of differential heating and geometry. When the sun rises, it is the tops of the mountain peaks which receive first light, and as the day progresses, the mountain slopes take on a greater heat load than the valleys. This results in a temperature inequity between the two, and as warm air rises off the slopes, cool air moves up out of the valleys to replace it. This upslope wind is called a *valley breeze*. The opposite effect takes place in the afternoon, as the valley radiates heat. The peaks, long since cooled, transport air into the valley in a process that is partly gravitational and partly convective and is called a *mountain breeze*.

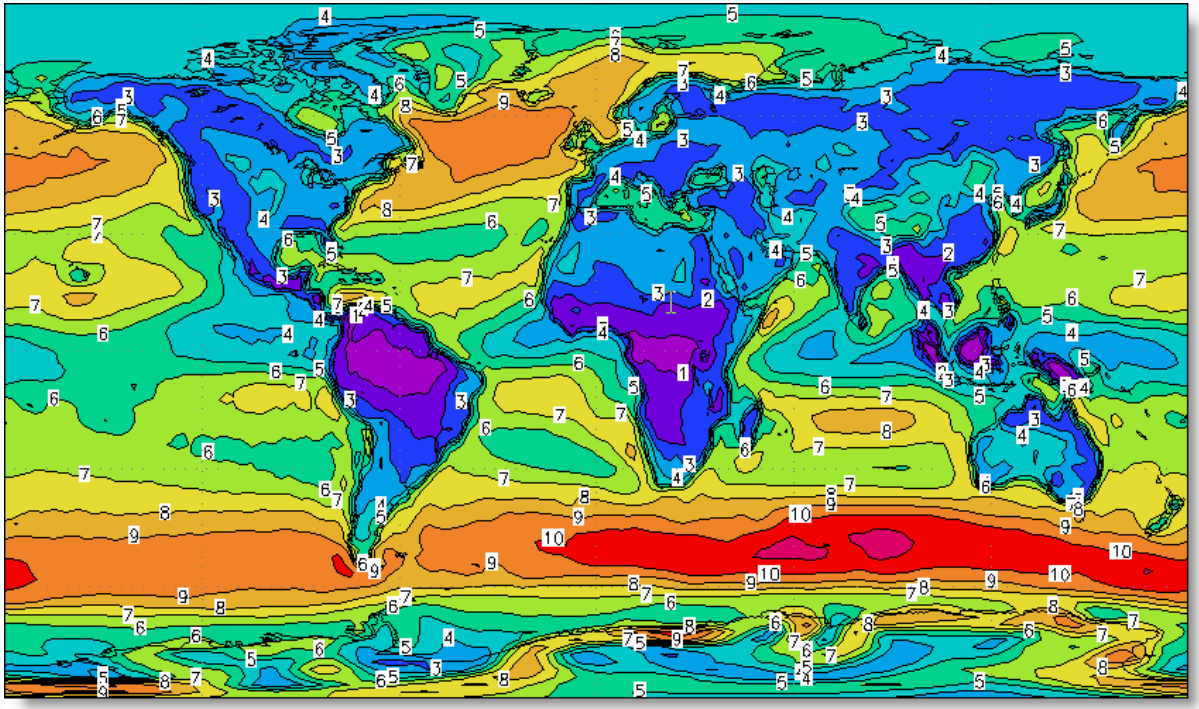

**Fig.4.** The map shows the mean wind speed in ms$^{-1}$ @ 10 m a.g.l. for the period 1976-95, according to the NCEP/NCAR reanalysis data set. Even though the wind climate is known in broad outline for most of the world, more detailed and reliable information is usually required for wind resource assessments. This section lists a number of wind atlases and other wind investigations for different countries around the world; many of which contains measured and/or modeled data sets for application in wind resource assessment. 73 kb. (withdraw because article case is >1Mb)

Mountain breezes are one example of what is known more generally as a katabatic wind. These are winds driven by cold air flowing down a slope, and occur on the largest scale in Greenland and Antarctica...Most often, this term refers to winds which form when air which has cooled over a high, cold plateau is set in motion and descends under the influence of gravity. Winds of this type are common in regions of Mongolia and in glaciated locations.

Because *katabatic* refers specifically to the vertical motion of the wind, this group also includes winds which form on the lee side of mountains, and heat as a consequence of compression. Such winds may undergo a temperature increase of 20 °C (36 °F) or more, and many of the world's "named" winds (see list below) belong to this group. Among the most well-known of these winds are the chinook of Western Canada and the American Northwest, the Swiss föhn, California's infamous Santa Ana wind, and the French Mistral. The opposite of a katabatic wind is an anabatic wind, or an upward-moving wind. The above-described *valley breeze* is an anabatic wind. A widely-used term, though one not formally recognised by meteorologists, is *orographic wind*. This refers to air which undergoes orographic lifting. Most often, this is in the context of winds such as the chinook or the föhn, which undergo lifting by mountain ranges before descending and warming on the lee side.

A **sea-breeze** (or **onshore breeze**) is a wind from the sea that develops over land near coasts. It is formed by increasing temperature differences between the land and water which create a pressure minimum over the land due to its relative warmth and forces higher pressure, cooler air from the sea to move inland.

**Fig.5.** A lenticular cloud in New Mexico. (withdraw because article size is >1 Mb).

**Land breezes.** At night, the land cools off quicker than the ocean due to differences in their specific heat values, which forces the dying of the daytime sea breeze. If the land cools below that of the adjacent sea surface temperature, the pressure over the water will be lower than that of the land, setting up a land breeze as long as the environmental surface wind pattern is not strong enough to oppose it. If there is sufficient moisture and instability available, the land breeze can cause showers or even thunderstorms, over the water. Overnight thunderstorm development offshore can be a good predictor for the activity on land the following day, as long as there are no expected changes to the weather pattern over the following 12-24 hours. The land breeze will die once the land warms up again the next morning.

**6. Mountains**. A mountain is a landform that extends above the surrounding terrain in a limited area. A mountain is generally steeper than a **hill**, but there is no universally accepted standard definition for the height of a mountain or a hill although a mountain usually has an identifiable summit. Mountains cover 54% of Asia, 36% of North America, 25% of Europe, 22% of South America, 17% of Australia, and 3% of Africa. As a whole, 24% of the Earth's land mass is mountainous. 10% of people live in mountainous regions. Most of the world's rivers are fed from mountain sources, and more than half of humanity depends on mountains for water.

Sufficiently tall mountains have very different climatic conditions at the top than at the base, and will thus have different life zones at different altitudes. The flora and fauna found in these zones tend to become isolated since the conditions above and below a particular zone will be inhospitable to those organisms. These isolated ecological systems are known as sky islands and/or microclimates. Tree forests are forests on mountain sides which attract moisture from the trees, creating a unique ecosystem. Very tall mountains may be covered in ice or snow.

Air as high as a mountain is poorly warmed and, therefore, cold. Air temperature normally drops 1 to 2 degrees Celsius (1.8 to 3.6 degrees Fahrenheit) for each 300 meters (1000 feet) of altitude.

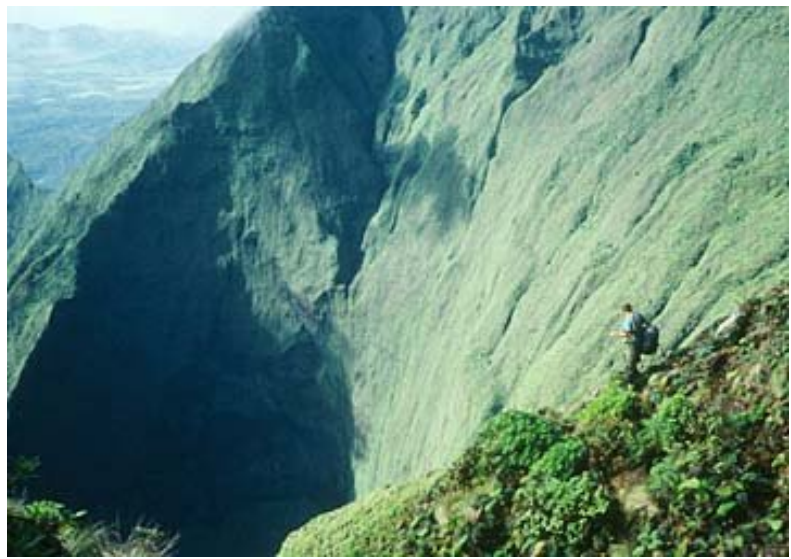

**Fig.6.** At the famous "Blue Hole," Mount Waialeale drops off precipitously. 48 kb

**7. Humidity**. The term humidity is usually taken in daily language to refer to relative humidity. Relative humidity is defined as the amount of water vapor in a sample of air compared to the maximum amount of water vapor the air can hold at any specific temperature. Humidity may also be expressed as Absolute humidity and specific humidity. Relative humidity is an important metric used in forecasting weather. Humidity indicates the likelihood of precipitation, dew, or fog. High humidity makes people feel hotter outside in the summer because it reduces the effectiveness of sweating to cool the body by preventing the evaporation of perspiration from the skin. This effect is calculated in a heat index table. Warm water vapor has more thermal energy than cool water vapor and therefore more of it evaporates into warm air than into cold air.

**8. A desert** is a landscape form or region that receives very little precipitation. Deserts are defined as areas that receive an average annual precipitation of less than 250 mm (10 in). In the Köppen climate classification system, deserts are classed as (BW).

Deserts take up one-third of the Earth's land surface. They usually have a large diurnal and seasonal temperature range, with high daytime temperatures (in summer up to 45 °C or 113 °F), and low night-time temperatures (in winter down to 0 °C; 32 °F) due to extremely low humidity. Water acts to trap infrared radiation from both the sun and the ground, and dry desert air is incapable of blocking sunlight during the day or trapping heat during the night. Thus during daylight all of the sun's heat reaches the ground. As soon as the sun sets the desert cools quickly by radiating its heat into space. Urban areas in deserts lack large (more than 25 °F/14 °C) daily temperature ranges, partially due to the urban heat island effect.

Many deserts are shielded in rain by rain shadows, mountains blocking the path of precipitation to the desert. Deserts are often composed of sand and rocky surfaces. Sand dunes called ergs and stony surfaces called hamada surfaces compose a minority of desert surfaces. Exposures of rocky terrain are typical, and reflect minimal soil development and sparseness of vegetation.

Bottomlands may be salt-covered flats. Eolian processes are major factors in shaping desert landscapes. **Cold deserts** (also known as polar deserts) have similar features but the main form of precipitation is snow rather than rain. Antarctica is the world's largest cold desert (composed of about 98 percent thick continental ice sheet and 2 percent barren rock). The largest hot desert is the Sahara. Deserts sometimes contain valuable mineral deposits that were formed in the arid environment or that were exposed by erosion.

Rain *does* fall occasionally in deserts, and desert storms are often violent. A record 44 millimeters (1.7 in) of rain once fell within 3 hours in the Sahara. Large Saharan storms may deliver up to 1 millimeter per minute. Normally dry stream channels, called arroyos or wadis, can quickly fill after heavy rains, and flash floods make these channels dangerous.

Though little rain falls in deserts, deserts receive runoff from ephemeral, or short-lived, streams fed considerable quantities of sediment for a day or two. Although most deserts are in basins with closed or interior drainage, a few deserts are crossed by 'exotic' rivers that derive their water from outside the desert. Such rivers infiltrate soils and evaporate large amounts of water on their journeys through the deserts, but their volumes are such that they maintain their continuity. The Nile River, the Colorado River, and the Yellow River are exotic rivers that flow through deserts to deliver their sediments to the sea. Deserts may also have underground springs, rivers, or reservoirs that lay close to the surface, or deep underground. Plants that have not completely adapted to sporadic rainfalls in a desert environment may tap into underground water sources that do not exceed the reach of their root systems.

Lakes form where rainfall or meltwater in interior drainage basins is sufficient. Desert lakes are generally shallow, temporary, and salty. Because these lakes are shallow and have a low bottom gradient, wind stress may cause the lake waters to move over many square kilometers. When small lakes dry up, they leave a salt crust or hardpan. The flat area of clay, silt, or sand encrusted with salt that forms is known as a playa. There are more than a hundred playas in North American deserts. Most are relics of large lakes that existed during the last ice age about 12,000 years ago. Lake Bonneville was a 52,000 kilometers² (20,000 mi²) lake almost 300 meters (1000 ft) deep in Utah, Nevada, and Idaho during the Ice Age. Today the remnants of Lake Bonneville include Utah's Great Salt Lake, Utah Lake, and Sevier Lake. Because playas are arid landforms from a wetter past, they contain useful clues to climatic change.

When the occasional precipitation does occur, it erodes the desert rocks quickly and powerfully. Winds are the other factor that erodes deserts—they are slow yet constant.

A desert is a hostile, potentially deadly environment for unprepared humans. The high heat causes rapid loss of water due to sweating, which can result in dehydration and death within days. In addition, unprotected humans are also at risk from heatstroke and venomous animals. Despite this, some cultures have made deserts their home for thousands of years, including the Bedouin, Touareg and Puebloan people. Modern technology, including advanced irrigation systems, desalinization and air conditioning have made deserts much more hospitable. In the United States and Israel, desert farming has found extensive use.

The Great Sandy Desert has nearly all its rain during from monsoonal thunderstorms or the occasional tropical cyclone rain depression. Thunderstorm days average 20-30 annually through most of the area (Burbidge 1983) although the desert has fairly high precipitation rates due to the high rates of evaporation this area remains an arid environment with vast areas of sands.

Other areas of the world, which see these rare precipitation events in drylands, are Northwest Mexico, South West America, and South West Asia. In North America in the Sonoran and Chihuahuan desert have received some tropical rainfall in the last 10 years. Tropical activity is rare in all deserts but what rain does arrive here is important to the delicate ecosystem existing.

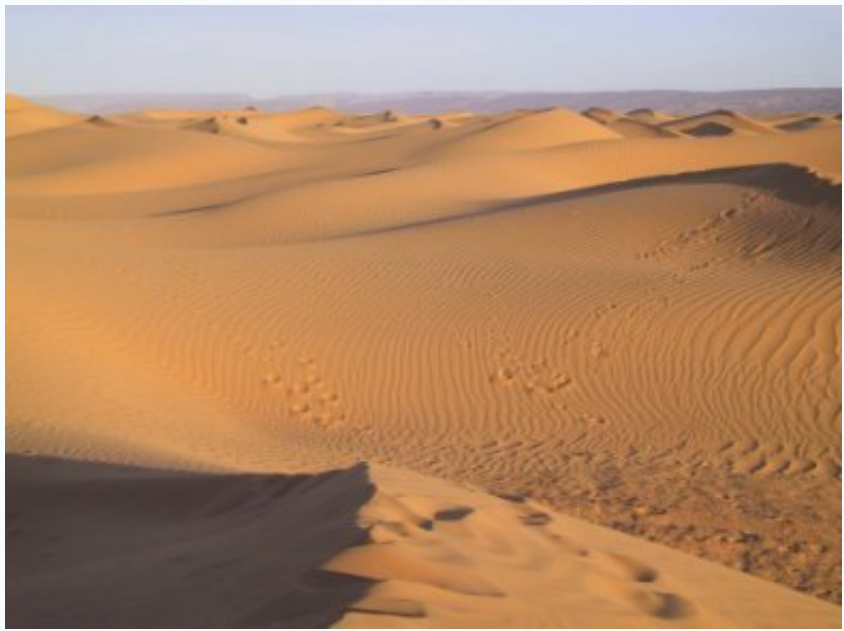

**Fig.7.** Typical desert. 18 kb.

**9. Aridity.** In general terms, the climate of a locale or region is said to be **arid** when it is characterized by a severe lack of available water, to the extent of hindering or even preventing the growth and development of plant and animal life. As a result, environments subject to arid climates tend to lack vegetation and are called xeric or desertic.

The expression 'available water' refers to water in the soil in excess to the wilting point. The air over a hot desert may actually contain substantial amounts of water vapor but that water may not be generally accessible to plants, except for very specialized organisms (such as some species of lichen). 'Lack of water' refers to use by plants. The water that is actually present in the environment may be sufficient for some species or usages (such as climax vegetation), and grossly insufficient for others. **Aridity**, the characteristic nature of arid climates, may thus depend on the use of the land. Regards to the presence of life, what is more important than the degree of rainfall is the fraction of precipitation that is not quickly lost through evaporation or runoff. Attempts to quantitatively describe the degree of aridity of a place has often led to the development of aridity indexes. There is no universal agreement on the precise boundaries between classes such as 'hyper-arid', 'arid', 'semi-arid', etc.

If different classification schemes and maps differ in their details, there is a general agreement about the fact that large areas of the Earth are considered arid. These include the hot deserts located broadly in sub-tropical regions, where the accumulation of water is largely prevented by either low precipitations, or high evaporation, or both, as well as cold deserts near the poles, where water may be permanently locked in solid forms (snow and ice). Other arid regions include areas located in the rain shadows of major mountain ranges or along coastal regions affected by significant upwelling (such as the Atacama Desert).

The distribution of aridity observed at any one point in time is largely the result of the general circulation of the atmosphere. The latter does change significantly over time through climate change. In addition, changes in land use can result in greater demands on soil water and induce a higher degree of aridity.

  **10**. **Control of local weather**. Governments spend billions of dollars merely studying the weather. The many big government research scientific organizations and perhaps a hundred thousand of scientists have been studying Earth's weather for more than a hundred years. There are gigantic numbers of scientific works about weather control. Most of them are impractical. We cannot exactly predict weather at long period, to avert a rain, strong wind, storm, tornado, or hurricane. We cannot control the clouds, temperature and humidity of the atmosphere, nor the power of rain. We cannot make more tolerable a winter or summer. We cannot convert a cold region to subtropics, a desolate wilderness to a prosperous region. We can only observe the storms and hurricanes and approximately predict their direction of movement. It is as if all the police department did was announce which neighborhoods were infested with killers and best avoided! Every year terrible storms, hurricanes, strong winds and rains and inundations destroy thousands of houses, kill thousands of men.

## INNOVATIONS AND DESCRIPTION

  **1. Idea and Innovations**. The idea here is creating a cheap range of inflatable 'mountains' (really immense gasbags) (Figs.8, 9) from a thin film whose presence forces humid air (a wet wind) to rise to high altitude. It is well known that air expands and cools at altitude. The air humidity decreases, exceeds the maximal saturation level and superfluous water vapor condenses in various forms, including rain or rain clouds.

  The top of the gasbags' film is located at an altitude of ~3 - 5 km. It is supported at this altitude by a small additional air pressure produced by ground ventilators. The film is connected to Earth's ground by controlled cables, which allow some change in the height and orientation of the artificial mountain(s). The gasbag's external surface may require double-layer film. We can control the heat conductivity of the dome cover by pumping an air between two layers of the dome cover and change the solar heating (solar radiation) by control of cover clarity or pumping a warm air between layers if icing-over or show is at the dome top. That allows selecting for different conditions (solar heating) in the covered area and by pumping air into the dome.

Special film may be used in a more complex dome design if we want to have finely control conditions inside the AB-Dome [1]. Envisioned is a cheap film having liquid crystal and conducting layers. The clarity is controlled by electric voltage. These layers, by selective control, can pass or blockade the solar light (or parts of solar spectrum) and pass or blockade the Earth's radiation. The incoming and outgoing radiations have different wavelengths. That makes control of them separately feasible and therefore possible to manage the heating or cooling of the Earth's surface under this film. In conventional conditions about 50% of the solar energy reaches the Earth surface. Much is reflected back to outer space by the white clouds. In our closed water system [1] the rain (or at least condensation) will occur at night when the temperature is low. In open atmosphere, the Sun heats the ground; the ground must heat the whole troposphere (6 – 10 km) before stable temperature rises are achieved. In our case the ground heats ONLY the air into the dome (as in a hotbed). We have a literal greenhouse effect, for the 'overroof' prevents the hot air escaping. That means that many cold regions (Alaska, Siberia, North Canada) may absorb more solar energy and became a temperate climate or sub-tropic climate (under the dome, as far as plants are concerned). That also means, by increasing the albedo of the gas bags, that the Sahara and other deserts can be a prosperous area with a fine growing and living climate and with a closed-loop water cycle.

The building of a film dome is very easy. We spread out the collapsed film over Earth's surface, turn on the pumping propellers and the film is raised by air overpressure to the needed altitude limited by the support cables. Damage to the film is not a major trouble because the additional air pressure is very small (0.005 - 0.05 atm) and air leakage is compensated for by the air impellers. Unlike in a space colony or planetary colony, the outside air is friendly and at worst we lose some heat (or cold) and water vapor.

**The first main innovation** of the offered AB-Mountain (and main difference from a conventional hotbed, or greenhouse) is the inflatable HIGH span of the closed cover (up to 3 – 5 km). The great vertical scale height of the enclosed volume aids organizing the rise of humid air at high altitude. The air is cooling and in that process, producing a lot of water which may be collected by rain channels on the AB-Mountain at high altitude and make much energy in hydro-electric turbines located on Earth's surface as the waters descend. This freshwater may be used for irrigation or sale [1]. The rest of the water will be precipitating as rain in near regions and can change a dry environment for a given region to a subtropical climate (the size of the region depends upon the size of the offered artificial mountain range).

The other profit from the offered AB-Domes (which make up the mountain range) is a closed loop water cycle, which prevents escape of water vaporized by plants; and returns this water in the nighttime when the air temperature decreases and internal condensation recurs. That allows us to perform irrigation in the gigantic portion of Earth's land area that does not have enough water for agriculture. We can convert the desert and desolate wildernesses into gardens without expensive delivery of remote freshwater. The initial amount of water for water cycle may be collected from atmospheric precipitation in some period or delivered. Only losses and exports of final products, typically one part in a thousand of all the water required during growing, need be replaced. Good soil is not a necessity-- hydroponics allows us to achieve record harvests on any soil.

**The second important innovation** is control of dome height. That allows reaching a maximum efficiency of water extraction and effective 'control' of regional weather.

**The third innovation** is using the high altitude of collected water for production of hydroelectric energy.

**The fourth important innovation** is the use of cheap thin film for building of mountains. This innovation decreases the construction cost by many millions of times in comparison to make the artificial mountains from soil and stones! Also, in case of improved climate modeling, when we find that moving the 'mountain range' just a little bit over would be a vast improvement—the very subject is not impossible even to consider.

The fifth innovation is using a cheap controlled heat conductivity, double-layer cover (controlled clarity is optionally needed for some regions). This innovation allows conserving solar heat (in cold regions), to control the temperature in the dome (in hot climates). That allows the user to get two to three rich crops annually in middle latitudes and to convert the cold zones (Siberia, North Canada, Alaska, etc.) to good single-crop regions.

The sixth innovation is using the cool water from the artificial mountain for cooling of buildings, crops, condensers, etc.

The seventh innovation is using the artificial mountain for high altitude windmills having high power (tens kW/m).

The eighth innovation is using the high water pressure (up 300 - 500 atm) for delivery of freshwater over long distances by tubes without pump stations.

The ninth innovation is control of cover reflectivity which allows to influence to climate of outer region.

Lest it be objected that such domes would take impractical amounts of plastic, consider that the world's plastic production is today on the order of 100 million tons. If, with economic growth, this amount doubles over the next generation and the increase is used for doming over territory, at 300-500 tons a square kilometer 200,000 square kilometers could be roofed over annually. While small in comparison to the approximately 150 million square kilometers of land area, consider that 200,000 1 kilometer sites scattered over the face of the Earth newly made habitable could revitalize vast swaths of land surrounding them—one square kilometer could grow local vegetables for a city in the desert, one over there could grow biofuel, enabling a desolate South Atlantic island to become fuel independent; at first, easily a billion people a year could be taken out of sweltering heat, biting cold and slashing rains, saving the money buying and running heating and air conditioning equipment would require.

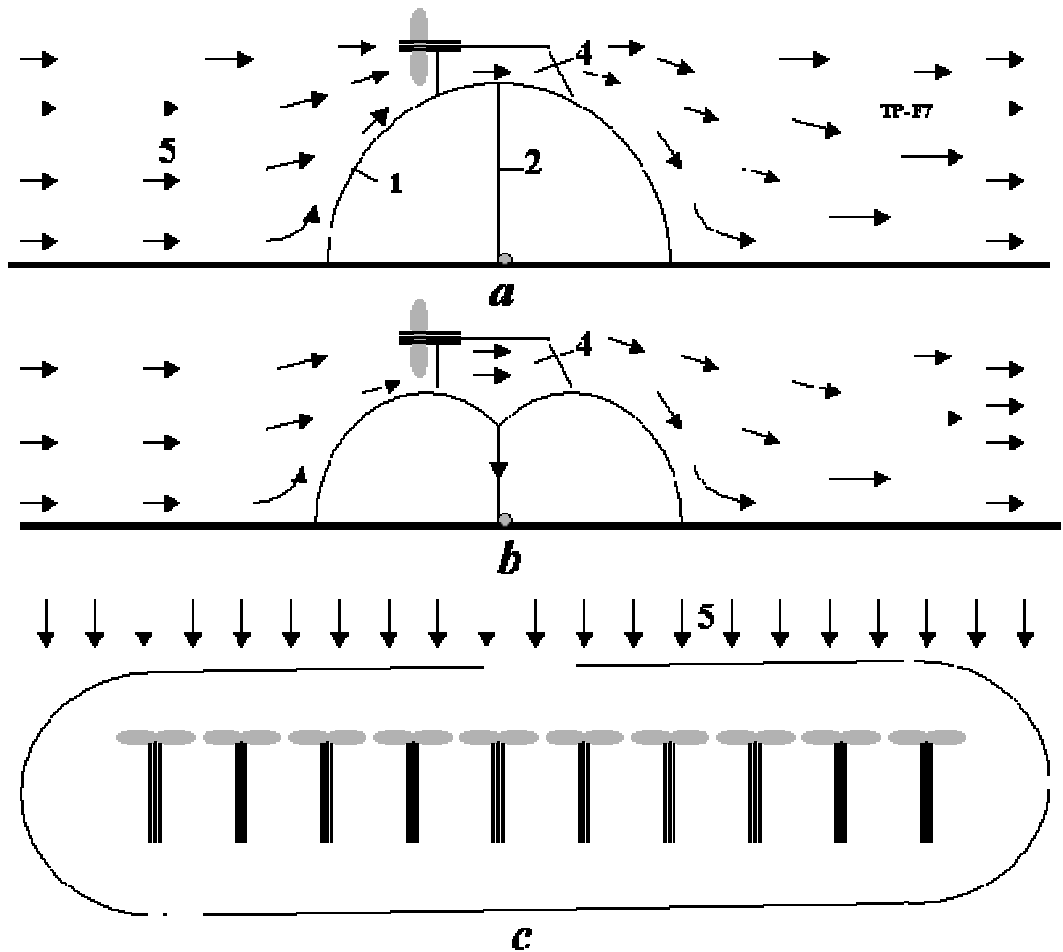

**Fig. 8**. Artificial Inflatable Mountain. (*a*) Cross-section of mountain in position maximal altitude; (*b*) Cross-section of mountain in position decreased altitude; (*c*) Top view of cylindrical mountain ridge. Notations:

1 – Inflatable Mountain AB-Dome; 2 – cable of height control; 4 – water collector; 5 – wind. 32 kb.

Our design of the mountain-dome is presented in Figs. 8, 8a, 9 that include the thin inflated film mountain-dome. The innovations are listed here: (1) the construction is air-inflatable; (2) each mountain-dome is fabricated with very thin, transparent film (thickness is 1 to 5 mm) having controlled clarity and controlled heat conductivity without rigid supports; (3) the enclosing film has two conductivity layers plus a liquid crystal layer between them which changes its clarity, color and reflectivity under an electric voltage (fig. 10); (4) the bound section of dome has a semi-cylindrical form (figs. 8, 8a, 9). The air pressure is more in these sections and they protect the central sections from the outer wind.

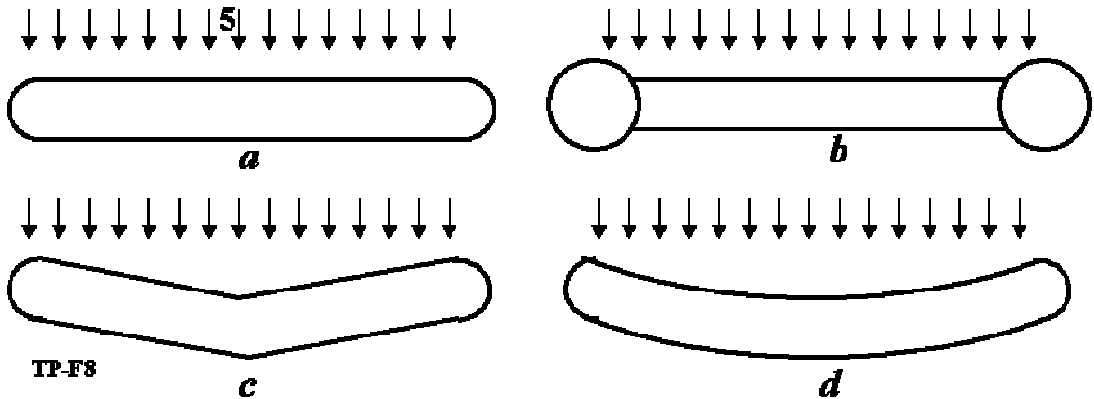

**Fig. 8a.** Form of artificial mountains ridge: (*a*) straight; (*b*) Straight with spherical end; (*c*) Angle; (*d*) Concave. 5 – wind.

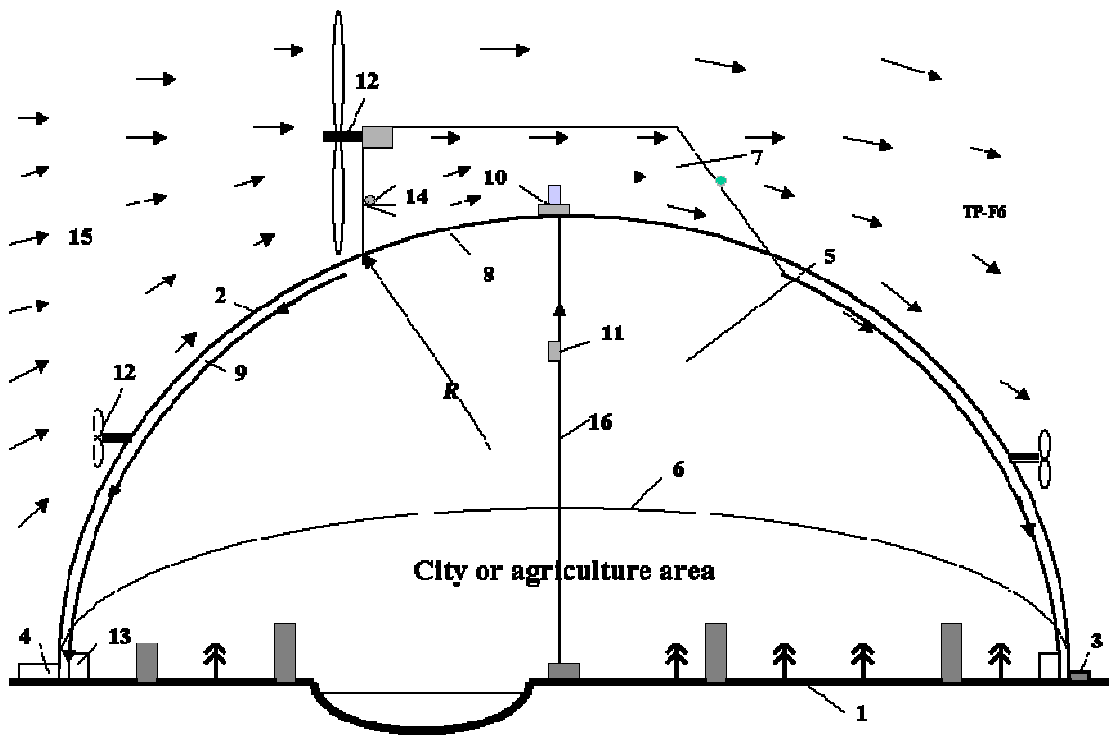

**Fig.9.** Cross-section of Mountain AB-Dome used for big city or agriculture area; *Notations*: 1 – covered area; 2 – thin film; 3 – ventilator (air pump); 4 – exit; 5 –semi-cylindrical thin film AB-Dome; 6 – control reflectivity thin film (optional); 7 –collector of water; 8 – strong cables; 9 – water tube; 10 –rooftop utilities and businesses: TV, communication, telescope, locator service (differential GPS), tourists; 11 – elevator; 12 – windmills; 13 water turbines and electric generator; 14 – injector of water collector; 15 – wind; 16 – height control cable.

Fig. 10 illustrates the thin transparent control dome cover we envision. The inflated textile shell—technical "textiles" can be woven or non-woven (films)—embodies the innovations listed: (1) the film is very thin, approximately 0.1 to 5 mm. A film this thin has never before been used in a major building; (2) the film has two strong nets, with a mesh of about $0.1 \times 0.1$ m and $a = 1 \times 1$ m, the threads are about additional 0.5 mm for a small mesh and about 1 mm for a big mesh. The net prevents the watertight and airtight film covering from being damaged by vibration; (3) the film incorporates (optional) a tiny electrically conductive wire net with a mesh about $0.1 \times 0.1$ m and a line width of about 100 μ (microns) and a thickness near 10 μ. The wire net is electric (voltage) control conductor. It can inform the dome maintenance engineers concerning the place and size of film damage (tears, rips, etc.); (4) the film may be twin-layered with the gap — $c = 1$ m and $b = 2$ m—between film layers for heat insulation. In polar (and hot) regions this multi-layered covering is the main means for heat isolation and puncture of one of the layers won't cause a loss of shape because the second film layer is unaffected by holing; (5) the airspace in the dome's covering can be partitioned, either hermetically or not; and (6) part of the covering can have a very thin shiny aluminum coating that is about 1μ (micron) for reflection of unnecessary solar radiation in equatorial or retention of additional solar radiation in the polar regions [1-16].

The author offers a method for moving off the snow and ice from the film at high altitude or in polar regions. After snowfall we decrease the heat cover protection, heating the show (ice) by warm air flowing into channels 5 (fig.10) (between cover layers), and water runs down in tubes 3 (fig.9).

The town cover may be used as a screen for projecting of pictures, films and advertising on the cover at night times.

**2. Brief information about cover film.** Our complex dome cover (film) may have 5 layers (fig. 10c): transparent dielectric layer, conducting layer (about 1 - 3 μ), liquid crystal layer (about 10 - 100 μ), conducting layer (for example, $SnO_2$), and transparant dielectric layer. Altogether thickness is ~0.1 - 0.5 mm. Control voltage is 5 - 10 V. This film may be produced by industry relatively cheaply.

Eikos Inc of Franklin, Massachusetts and Unidym Inc. of Silicon Valley, California are developing transparent, electrically conductive films of carbon nanotubes to replace indium tin oxide (ITO). Carbon nanotube films are substantially more mechanically robust than ITO films.

**3. Liquid crystals** (LC) are substances that exhibit a phase of matter that has properties between those of a conventional liquid, and those of a solid crystal.

Liquid crystals find wide use in liquid crystal displays (LCD), which rely on the optical properties of certain liquid crystalline molecules in the presence or absence of an electric field. The electric field can be used to make a pixel switch between clear or dark on command. Color LCD systems use the same technique, with color filters used to generate red, green, and blue pixels. Similar principles can be used to make other liquid crystal based optical devices. Liquid crystal in fluid form is used to detect electrically generated hot spots for failure analysis in the semiconductor industry. Liquid crystal memory units with extensive capacity were used in Space Shuttle navigation equipment. It is also worth noting that many common fluids are in fact liquid crystals. Soap, for instance, is a liquid crystal, and forms a variety of LC phases depending on its concentration in water.

The conventional control clarity (transparency) film reflected excess solar energy back to space. If practical in the future, film of a future generation may incorporate thin-film solar cells that convert the solar energy into electricity.

**4. Transparency**. In optics, transparency is the material property of allowing light to pass through. Though transparency usually refers to visible light in common usage, it may correctly be used to refer to any type of radiation. Examples of transparent materials are air and some other gases, liquids such as water, most glasses, and plastics such as Perspex and Pyrex. Where the degree of transparency varies according to the wavelength of the light. From electrodynamics it results that only a vacuum is really transparent in the strict meaning, any matter has a certain absorption for electromagnetic waves. There are transparent glass walls that can be made opaque by the application of an electric charge, a technology known as electrochromics.Certain crystals are transparent because there are straight lines through the crystal structure. Light passes unobstructed along these lines. There is a complicated theory "predicting" (calculating) absorption and its spectral dependence of different materials.

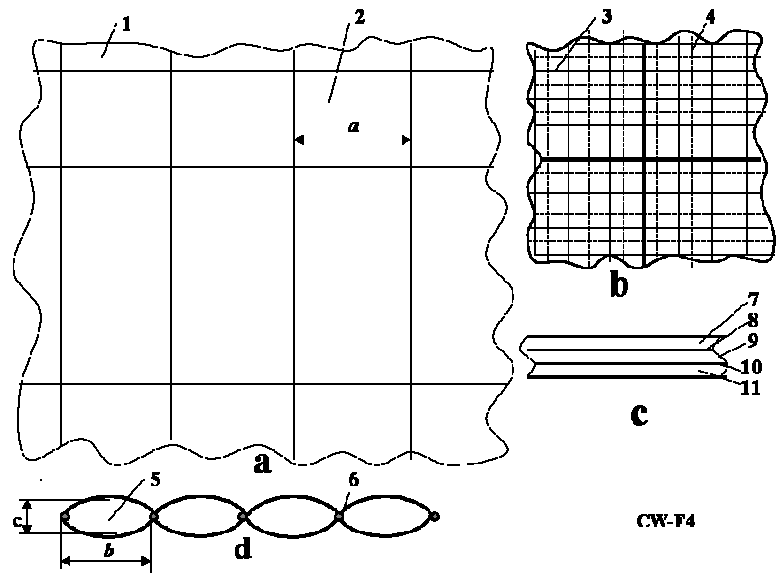

**Fig. 10.** Design of covering membrane. *Notations*: (a) Big fragment of cover with control clarity (reflectivity, carrying capacity) and heat conductivity; (b) Small fragment of cover; (c) Cross-section of cover (film) having 5 layers; (d) Longitudinal cross-section of cover for cold and hot regions; 1 - cover; 2 -mesh; 3 - small mesh; 4 - thin electric net; 5 - cell of cover; 6 - tubes;: 7 - transparent dielectric layer, 8 - conducting layer (about 1 - 3 μ), 9 - liquid crystal layer (about 10 - 100 μ), 10 - conducting layer, and 11 - transparant dielectric layer. Common thickness is 0.1 - 0.5 mm. Control voltage is 5 - 10 V.

## THEORY AND COMPUTATIONS of THE AB-DOME and MOUNTAIN SYSTEM

As wind flows over and around a fully exposed, nearly completely sealed inflated dome, the weather affecting the external film on the windward side must endure positive air pressures as the wind stagnates. Simultaneously, low air pressure eddies will be present on the leeward side of the dome. In other words, air pressure gradients caused by air density differences on different parts of the dome's envelope is characterized as the "buoyancy effect". The buoyancy effect will be greatest during the coldest weather when the dome is heated and the temperature difference between its interior and exterior are greatest. In extremely cold climates such as the Arctic and Antarctic Regions the buoyancy effect tends to dominate dome pressurization.

A reader can derive the equations below from well-known physical laws [17]. Therefore, the author does not give detailed explanations of these.

1. **Amount of water in atmosphere**. Amount of water in atmosphere depends upon temperature and humidity. For relative humidity 100%, the maximum partial pressure of water vapor is shown in Table 1.

**Table 1**. Maximum partial pressure of water vapor in atmosphere for given air temperature

| $t$, C | -10 | 0 | 10 | 20 | 30 | 40 | 50 | 60 | 70 | 80 | 90 | 100 |
|---|---|---|---|---|---|---|---|---|---|---|---|---|
| $p$,kPa | 0.287 | 0.611 | 1.22 | 2.33 | 4.27 | 7.33 | 12.3 | 19.9 | 30.9 | 49.7 | 70.1 | 101 |

The amount of water in 1 m$^3$ of air may be computed by equation

$$m_W = 0.00625[\,p(t_2)h - p(t_1)\,], \qquad (1)$$

where $m_W$ is mass of water, kg in 1 m$^3$ of air; $p(t)$ is vapor (steam) pressure from Table 1, kPa; $h = 0 \div 1$ is relative humidity; $t$ is temperature, C.

The computation of equation (1) is presented in fig.9. Typical relative humidity of atmosphere air is 0.5 - 1.

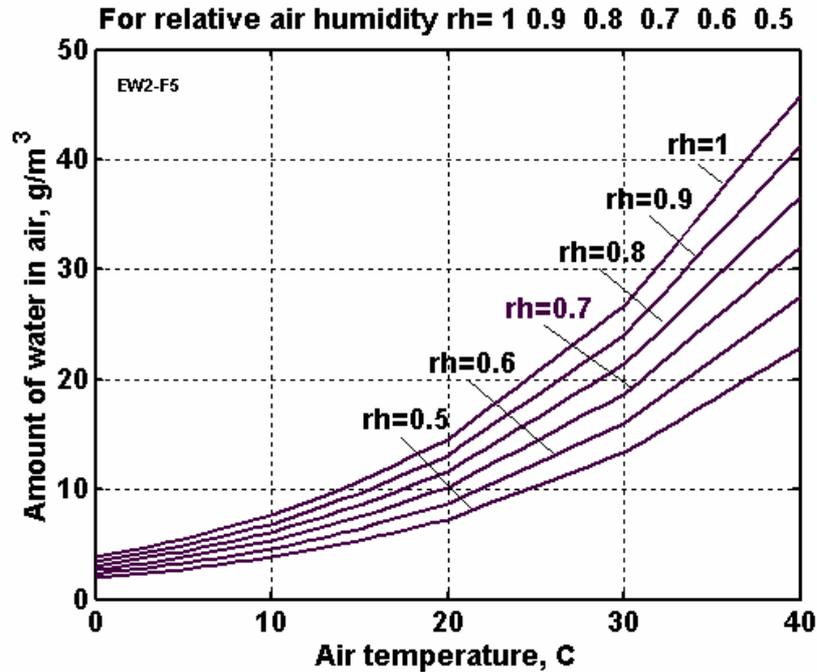

**Fig. 9**. Amount of water in 1 $m^3$ of air versus air temperature and relative humidity (rh). $H = 0$ km, $t_1 = 0$ °C. 36 kb.

## 2. Air temperature, density, and pressure versus altitude in troposphere.

Standard atmosphere is in Table 2.

**Table 2**. Standard atmosphere. $\rho_o = 1.225$ kg/$m^3$

| $H$, km | 0 | 0.4 | 1 | 2 | 3 | 4 | 5 | 6 |
|---|---|---|---|---|---|---|---|---|
| $t$, °K | 288.2 | 285.6 | 281.9 | 275.1 | 268.6 | 262.1 | 265.6 | 247.8 |
| $t$, °C | 15 | 12.4 | 8.5 | 2 | -4.5 | -11 | - 17.5 | -24 |
| $\rho/\rho_o$ | 0 | 0.907 | 0.887 | 0.822 | 0.742 | 0.669 | 0.601 | 0.538 |

Temperature, relative air density and pressure of troposphere (up 10 km) versus the altitude computed by equations:

$$T = T_0 - 0.0065H, \quad \overline{\rho} = \frac{\rho}{\rho_0} = \left(1 - \frac{H}{44300}\right)^{4.265}, \quad \overline{p} = \frac{p}{p_0} = \left(1 - \frac{H}{44300}\right)^{5.265}, \qquad (2)$$

where $T_0 = 15$ C, $\rho_o = 1.225$ kg/$m^3$, $p_o = 10^5$ N/$m^2$ are air temperature, density and pressure at sea level, $H = 0$; $H$ is altitude, m; $T$, $\rho$, $p$ are air temperature, density and pressure at altitude $H$, m. The computation of temperature via altitude are presented in figs. 10, 11.

After the encounter with the AB-Mountain range the atmospheric temperature is 5 – 7 C higher, because the air temperature increases when air pressure increases at low altitude. The analogy is to an air compressor; denser fractions are usually hotter. The water vapor is condensed and gives up its' heat to air. That additional air warming is important for countries which use the artificial AB-Mountains for protection from cold polar winds.

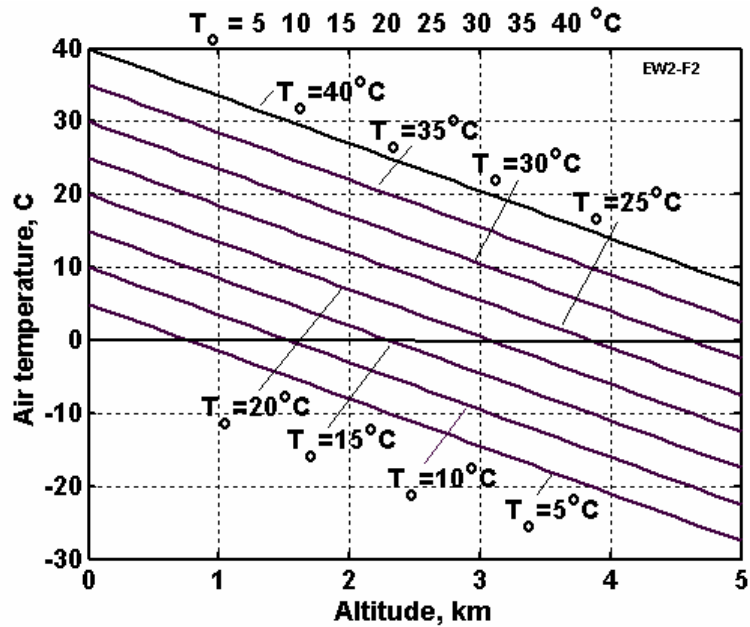

**Fig.10.** Air temperature versus the altitude for different temperatures.

2. **Altitude and wind speed**. Wind speed, *V*, increases with altitude, *H*, as follows

$$\overline{V} = V/V_0 = (H/H_0)^{\alpha},\tag{3}$$

where $\alpha = 0.1 - 0.25$ exponent coefficient depending upon surface roughness. When the surface is water, $\alpha = 0.1$; when surface is shrubs and woodlands $\alpha = 0.25$. The sub "0" means the data at Earth surface. The standard values for wind computation are $V_o = 6$ m/s, $H_o = 10$ m/s. The computation of this equation are presented in fig. 12. At high altitude the wind speed may be significantly more than equation (3) gives.

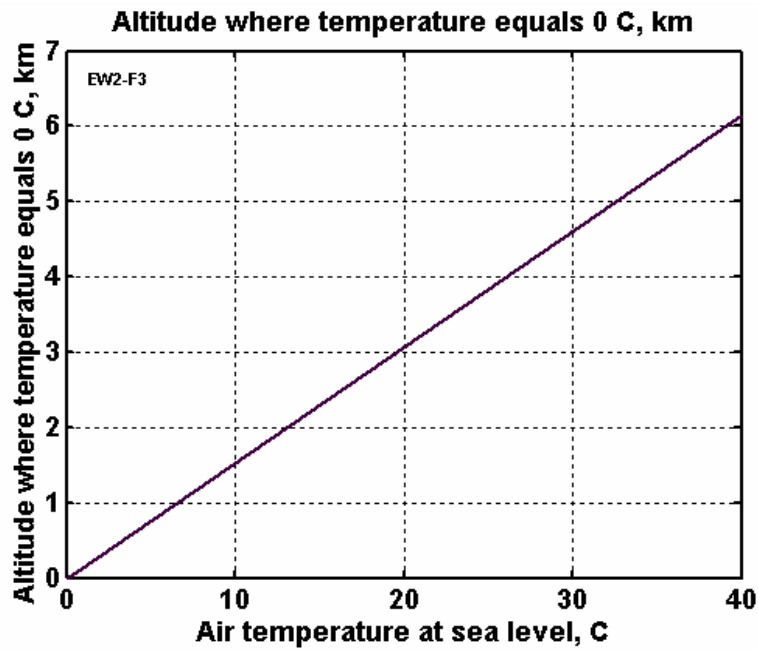

**Fig.11.** Altitude where temperature equals 0°C via air temperature at sea level.

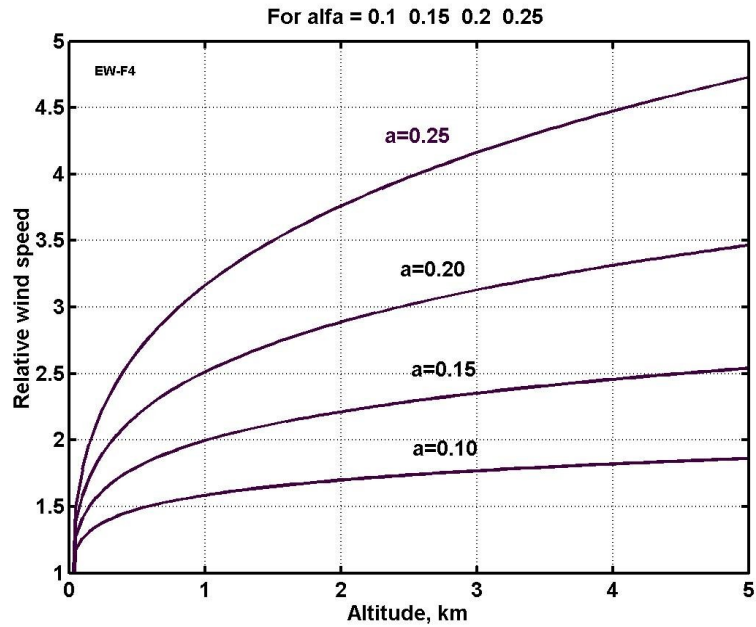

**Fig. 12.** Relative wind speed versus altitude for $V_o = 6$ m/s, $H_o = 10$ m/s.

**3. Water produced by AB-Mountains.** Each linear meter of the mountain ridge (which may stretch for 5 km, 100 km, or even more) produces the water

$$W(H_m) = k_c V_0 \int_0^{H_m} m_w(T, H_m, h)\overline{\rho}(H)\overline{V}(H)dH$$

where $m_w = 0.00625[p(T_0)h - p(H_m)]$, for $m_w > 0$, , (4)

if $m_w < 0$, than $m_w = 0$.

where $W$ is water flow produced by 1 m of mountain ridge, kg/s/m; $m_w$ is water in 1 m$^3$ [kg/m$^3$]; $H_m$ is maximal height of artificial mountain, m; $k_c = 0.5 \div 1$ is collector (extraction) coefficient; $h \approx 0.5 \div 1$ is air humidity; $V_0$ is wind speed at H = 0, m/s.

The computations of the equation (4) for $h = 0.3$, $k_c = 1$, $V_0 = 6$ m/s are presented in fig. 15.

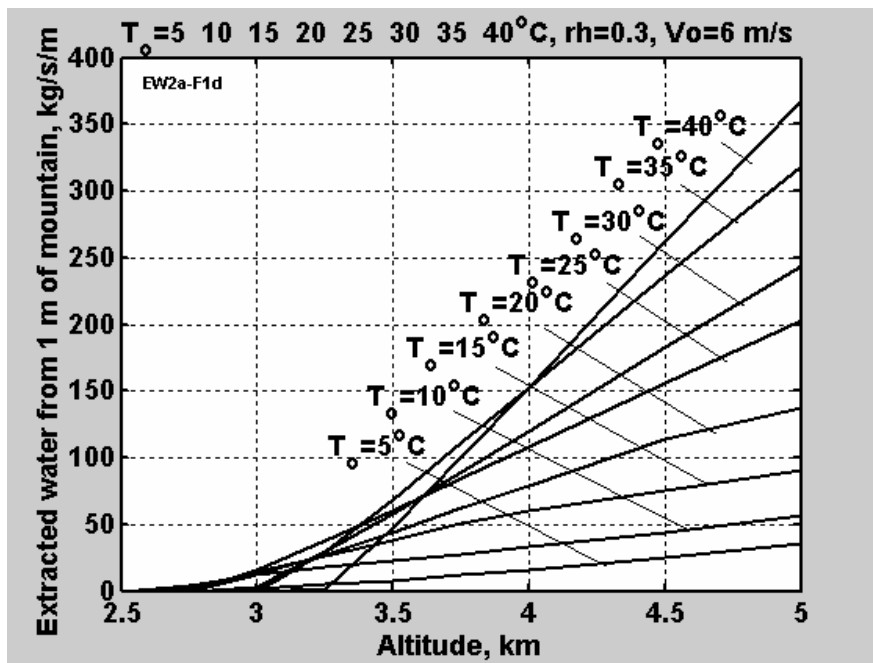

**Fig. 15**. Amount of water flow may be extracted by 1 m of the AB-Dome artificial Mountain via mountain height for different air temperature $T_o$ at $H_o = 10$ m , air speed $V_o = 6$ m/s, relative air humidity **rh = 0.3**  at sea level. Collector efficiency $k_c = 1$ , coefficient increasing of air speed with altitude $\alpha = 0.15$ ; $H_m = 5$ km.

  Please notice: If relative humidity is $rh = 0,3$, the mountain begins to produce water over altitude 3 km, when the $rh = 0.3$. Every 100 - 300m ridge length of a 3-5-km altitude artificial mountain can produce a good sized river. However, in reality the water production after the mountain height over $3.5 - 5$ km can not increase because all air water is condensed and precipitated. Many natural mountain over $3.5 - 5$ km is dry.

  The total water carried as vapor by atmospheric air is significantly more than we have here extracted. The perfect design of a water collector increases the extraction coefficient $k_c$. The computation of total water flow (for $k_c = 1$) is presented in fig.16. The water non-extracted from atmosphere goes as clouds and rain after crossing the mountain ridge.

**5. Energy produced by high altitude water.** The water condensed at high altitude has huge energy because it has great mass and is located advantageously at very high altitude. For example, if artificial mountain has a height = 5 km, the water pressure at sea level is about 500 atm. We can easy convert this energy into electricity by conventional hydro-turbine and electric generators.  And the higher the pressure, the smaller the installation needed.

  Equation is below:

$$P = g\eta W H_m , \qquad (5)$$

where $P$ is water power, W/m; $g = 9.81$ m/s²; $\eta$ is efficiency coefficient; $W$ is water flow, kg/s/m; (Eq. (4)); $H_m$ is maximal altitude, m. The result equation is in fig. 17.

  Again, please notice: That the artificial AB-Mountains produce gigantic levels of energy. A conventional large hydro-electric power station such as Hoover Dam may have about 2 GW of generating capacity. Only 4 km ridge length of 5 km tall artificial AB-Mountain can potentially produce energy comparable to the largest conventional hydroelectric stations in the world, in China and Brazil. (See for comparison http://en.wikipedia.org/wiki/List_of_the_largest_hydoelectric_power_stations) (And, most importantly, the AB-Mountain does not silt up its' reservoirs like a conventional dam does.) The huge amount of green water energy may be the most important profit potential obtainable from the offered AB-Mountain.

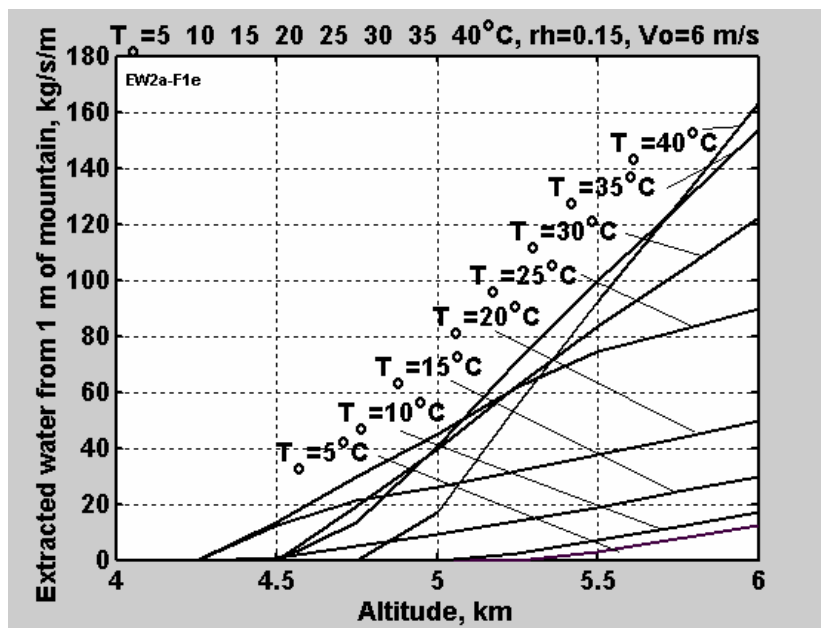

**Fig. 16**. Amount of water flow may be extracted by 1 m of the AB-Dome artificial Mountain via mountain height for different air temperature $T_o$ at $H_o = 10$ m , air speed $V_o = 6$ m/s, relative air humidity **rh = 0.15** at sea level. Collector efficiency $k_c = 1$ , coefficient increasing of air speed with altitude $\alpha = 0.15$ ; $H_m = 6$ km.

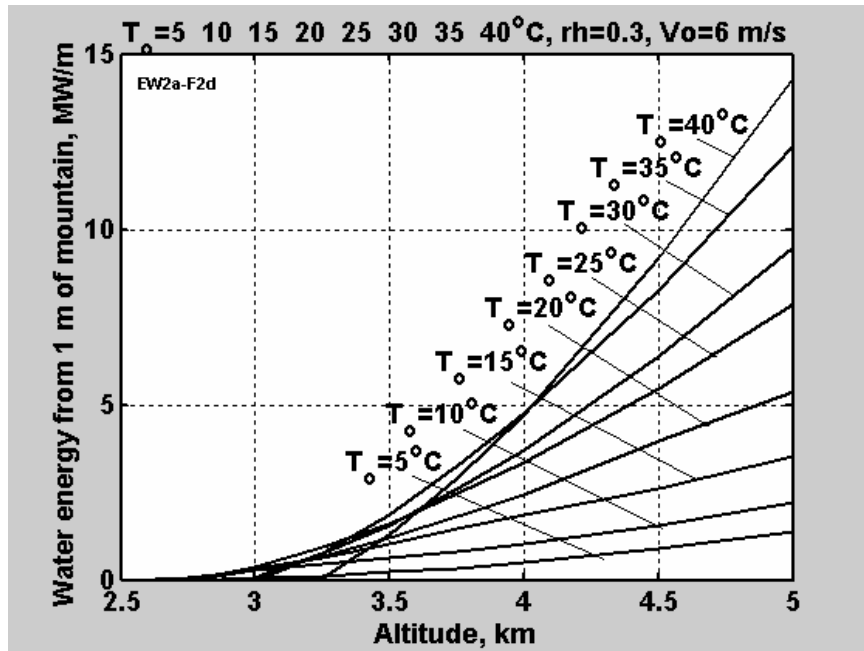

**Fig. 17.** Water energy (MW) from 1 m of AB-mountain via dome height and air temperature $T_o$ , air speed $V_o = 6$ m/s, relative air humidity **rh = 0.3** at sea level $H = 10$ m. Collector efficiency $k = 1$ , product of efficiency coefficients (tubes, hydro-turbine, electric generator) is $\eta = 0.8$, coefficient increasing of air speed with altitude $\alpha = 0.15$.

**6. Hot mountain–derived wind.** The water vapor condensing to water produces a lot of energy, about 2260 kJ/kg of water. The atmosphere absorbs this energy. The ground air temperature in the lee of the mountain is more than it is upwind of the mountain. The increase of air temperature may be estimated by equation received from equation of heat balance:

$$\Delta T = \frac{k_c \lambda \, m_{w0}(T_0) h}{c_p \rho_0}, \quad m_{w0} = 0.00625 \, p(T_0) , \qquad (6)$$

where $\Delta T$ is additional atmospheric temperature, C; $k_c \approx 0.7 - 1$ is extraction coefficient, $m_{w0}$ is amount of water in 1 m$^3$ of air at temperature $T_0$, kg/m$^3$ (see Table 1); $h \approx 0.5 - 1$ is humidity; $c_p \approx 1$ kJ/kg/C is average air heat capability; $\rho_0 = 1.225$ kg/m$^3$ is standard air density. Note, the artificial mountain works better than a natural mountain because one has a smooth surface and good, indeed, selectable aerodynamic form (semicylinder).

The computations for $k_c = 1$ are presented in fig.18.

As you see the artificial mountain can significantly increase temperatures of cold polar-derived winds. For example, if atmospheric temperature was 0 C, after an encounter with an artificial mountain it may be up to 7 C. If the initial air temperature was 10 C, after artificial mountain it may be up to 24 C. This effect is good relief for cold countries (Iceland-Scandinavia-Russia-Siberia, Canada, USA-Alaska), but it is not well for a hot area (Sahara). The cold water from AB-Mountain may be used for cooling the buildings of a city located behind the artificial Mountains. On the other hand, in the desert, we may place a new city before the mountains for wind cooling, and salt drying ponds behind the mountains, for free accelerated air-drying! Near the Red Sea, for example, 2.4 meters a year of salt water may evaporate from salt ponds. Such a rate could easily be doubled with hotter winds blowing for much of the time, doubling annual salt production.

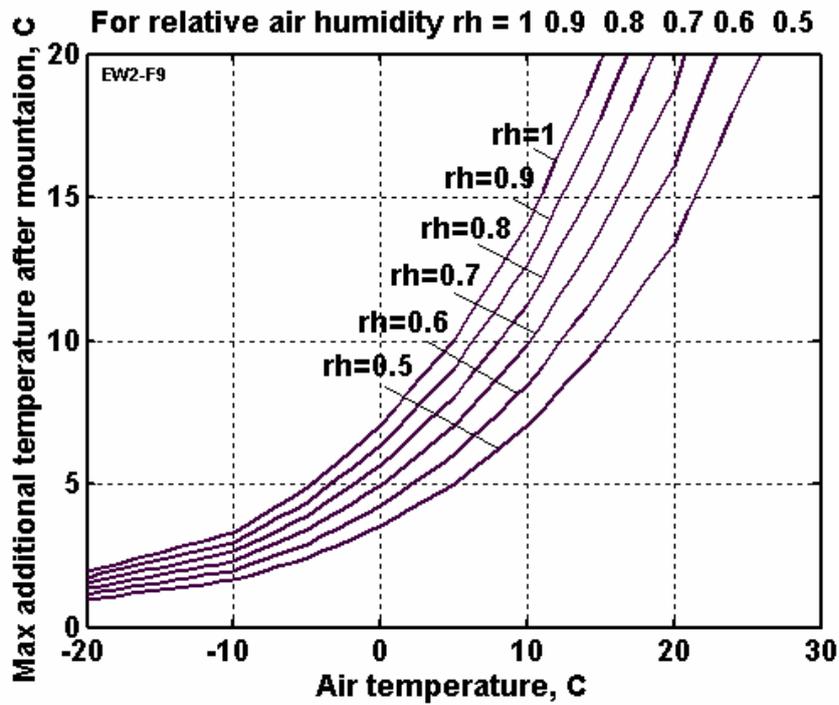

**Fig.18** . Maximal additional heating of atmosphere after mountain via temperature before mountain and relative humidity. Condenser coefficient equals 1.

Gigantic streams of energy flow through AB-Mountain ranges. If AB-Mountain length is 100 km only, water vapor power and air kinetic power is

$$P_v = c_{pm} SV\Delta T, \quad P_k = 0.5MV^2, \quad M = \rho VS, \qquad (7)$$

where $P_v$ is water vapor power, W; $P_k$ is air kinetic power, W; $c_{om}$ = 1.287 kJ/m³K for air is the heat coefficient; $V$ – average air speed, m/s; $M$ – air flow mass, kg/s; $S$ – cross-section of air flow, m²; $\rho$ – average air density, kg/m³. If Mountain ridge is only 100 km, $V$ = 5 m/s and $\Delta T$ = 10 C the water vapor power has gigantic value (1 - 3)×10⁷ MW (10,000 one gigawatt power electric stations). The kinetic air power is only $\approx 2 \times 10^4$ MW. (2 x 10 gigawatts)—which is logical, because water is about a thousand times denser than air! But look at those figures—the whole energy use of mankind today is only ~13,000 gigawatts! So a AB-Mountain range can actually generate the energy that all mankind uses today, at an amazingly realistic cost (see below).

The control (of height, reflectivity, and position) of the AB-Mountain range manages this gigantic natural flow of energy and changes the weather and climate around the artificial mountain.

## 7. Wind energy.
The AB-Dome has a high crest where there is a strong permanent wind. If we install the windmills at the top of AB-Dome, we get energy. This energy may be estimated by equations

$$N_1 = 0.5\eta\rho(H)D \cdot (\overline{V}V_0)^3, \qquad (8)$$

where $N_1$ is windmill power from 1 m of the mountain ridge, W/m; $\eta = 0.3 \div 0.6$ is coefficient efficiency of wind rotor, $\rho$ is air density at altitude $H$; $D$ is rotor diameter, m; $V$ is wind speed (see early), m/s.

Computations are presented in fig. 19.
As you see (compare with fig. 17) the wind energy in 500 times is less then the water energy and no reasons to install the complex and expensive windmills at top of AB-Mountain. But they may be useful on the low AB-Dome, which doesn't produce the water.

## 8. Cooling of building in hot weather.
The water from Dome top has temperature about 0°C. That may be used for cooling of cities through cold fountains, cooling of buildings, dwelling, food storage in hot countries, even growing non-tropical crops in tropical countries! That can save much energy spent by conditions in summer time in hot weather. This energy may be estimated by expression:

$$O_1 = c_p W T_0 / \eta_c, \qquad (9)$$

where $Q_1$ is possible energy from 1 m of AB-Mountain, W/m; $c_p = 4.19$ kJ/kg K is water heat capability; $T_0$ is air temperature at $H = 0$, C; $W$ is water flow, kg/s/m; $\eta_c \approx 0.3$ is coefficient efficiency of condition.

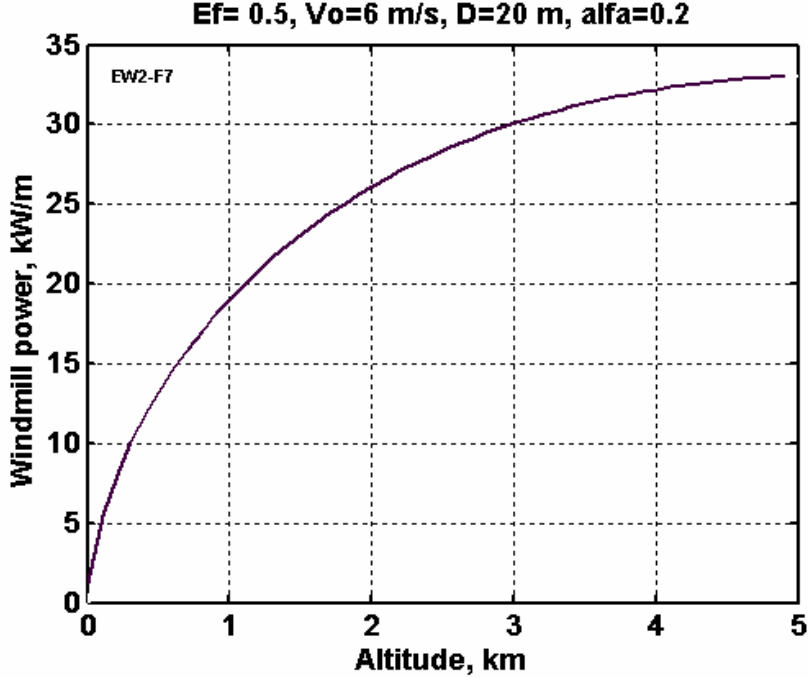

**Fig. 19.** Windmill power [kW/m] via mountain height for air speed $V_o = 6$ m/s at sea level $H_0 = 10$ m, product of efficiency coefficients (windmill, electric generator) is $\eta = 0.5$, coefficient increasing of air speed with altitude $\alpha = 0.2$, diameter of wind turbine $D = 20$ m.

*Example.* Let us to estimate the energy from cold water relative to the warm atmosphere. Assume the air temperature $T_a = 35$ C, the water temperature is 0 C, but while the water moves into delivery tubes, that warms approximately up $T_w = 5$ C. That means $T_0 = 35 - 5 = 30$ C. From fig. 15 we have the water flow $W = 300$ kg/m for $T = 35$ C. The total energy (for $\eta_c = 1$) is (Eq.(9)) $Q_1 = 38$ MW/m. That is *3* times more than the energy from high altitude water ($P = 12$ MW/m, fig.17). The efficiency of condition is $\eta_c \approx 0.3$ and a comfortable room temperature $T_r = 25$ C. If we substitute these values (difference between air temperature and room temperature, $T_0 = T_r - T_w = 25 - 5 = 20$ C) in Eq. (8), we may to save electric energy spent by air conditioners up to $Q_1 = 84$ MW/meter of mountain ridge. That is *7* times more then energy from high altitude water.

The cooling energy in some times more then $P$ [Eq. (4)]. But that is difficult to realize in practice (except for cooling needs) because the difference of temperatures between water and air is small ($\approx$ 20 C). The cold water for cooling of building may be delivered by current water (heat transfer) systems. An on-land version of OTEC technology using solar hot water heaters and the cold water to maximize the temperature differences might tap these energy streams, but why? The hydroelectricity is far cheaper and less capital intensive. One use of such masses of cold water might be to lower the temperature of bodies of water to better hold oxygen for productive aquaculture.

**9. The wind dynamic pressure** is computed by equation

$$p_d = \frac{\rho V^2}{2}, \qquad (10)$$

where $p_d$ is wind dynamic pressure, N/m²; $\rho$ is air density, for altitude $H = 0$ the $\rho = 1.225$ kg/m³; $V$ is wind speed, m/s. The computation is presented in fig.20.

The small overpressure of 0.01 atm forced into the AB-Dome or AB-Mountain to inflate it produces force $p = 1000$ N/m$^2$. That is greater than the dynamic pressure of very strong wind $V = 35$ m/s (126 km/hour). If it is necessary we can increase the internal pressure by some times if needed for very exceptional storms.

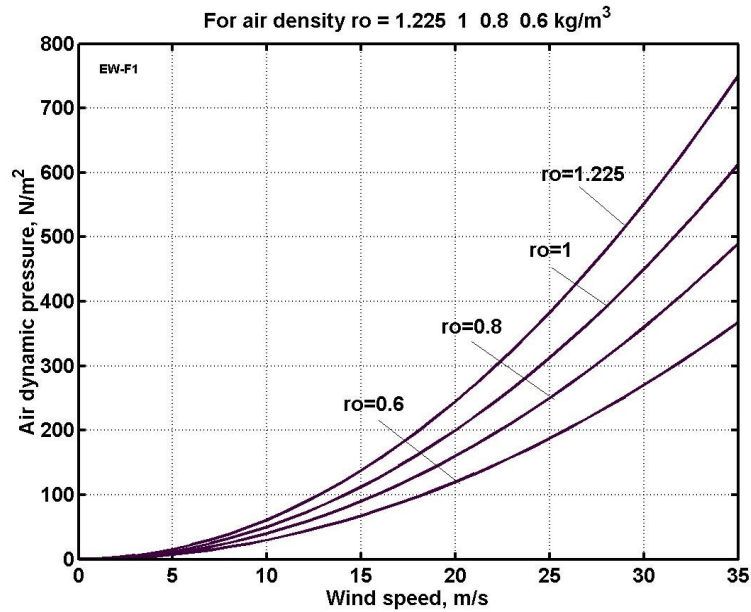

**Fig. 20.** Wind dynamic pressure versus wind speed and air density $\rho$. The $ro = 0.6$ is for $H \approx 6$ km.

**10. The thickness of the dome envelope**, its sheltering shell of film, is computed by formulas (from equation for tensile strength):

$$\delta_1 = \frac{Rp\bar{p}}{2\sigma}, \quad \delta_2 = \frac{Rp\bar{p}}{\sigma}, \tag{11}$$

where $\delta_1$ is the film thickness for a spherical dome, m; $\delta_2$ is the film thickness for a cylindrical dome, m; $R$ is radius of dome (or Dome cover between the support cable [1]), m; $p$ is additional pressure into the dome (10÷1000), N/m$^2$; $\sigma$ is safety tensile stress of film (up to $2 \times 10^9$), N/m$^2$.

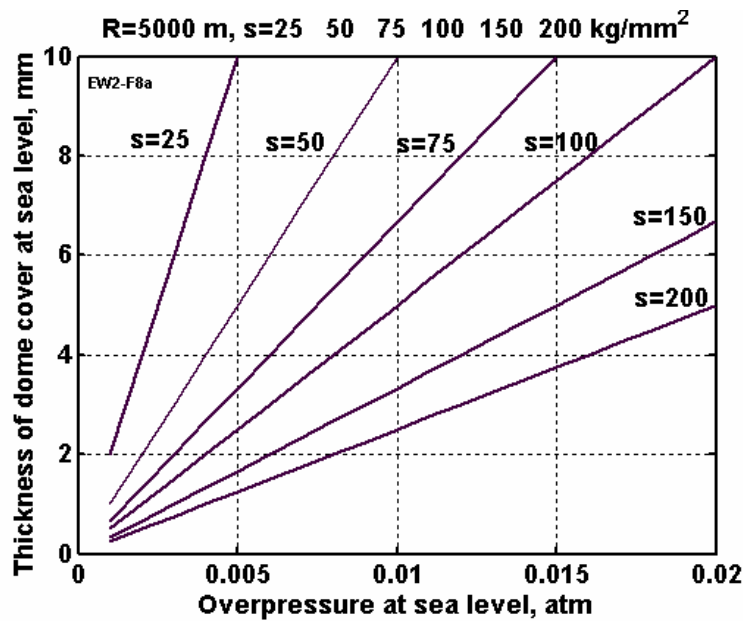

**Fig. 21.** Thickness of AB-Mountain cover (without the support cables) via over pressure at $H = 0$ for different safety stress. Radius of AB-Dome is $R = 5$ km.

**11. Cost of freshwater extractor**. The cost, C [\$/kL], of produced freshwater may be estimated by the equation:

$$C = \frac{C_i / l + M_e}{M_{wy}}, \qquad (12)$$

where $C_i$ is cost of installation; $l$ is live time of installation ($l \approx 10$), years; $M_e$ is annual mountains; $M_{wy}$ is annual amount of received freshwater, kL.

The retail cost of electricity for individual customers is \$0.18 per kWh at New York in 2007. Cost of other energy from other fuel is in [8] p.368. Average cost of water from river is \$0.49 - 1.09/kL in the USA, the water produced from sea costs about \$.5-2/kL in Israel.
The estimations are in Project section.

**12. Energy requested by different methods of desalination.** Below in Table 3 is some data about expense of energy for different methods.

**Table 3**. Estimation of energy expenses for different methods of freshwater extraction

| No | Method | Condition | Expense kJ/kL |
|----|--------|-----------|---------------|
| 1 | Evaporation | Expense only for evaporation* | $2.26 \times 10^6$ |
| 2 | Freezing | Expense only for freezing, c.e. $\eta = 0.3$ | $1 \times 10^6$ |
| 3 | Reverse osmosis | Expense only for pumping, $40 \div 70$ atm | $(4 \div 7) \times 10^3$ |

**\*** This expense may be decreased by 2 -3 times when the water installation is connected with heat or nuclear electric station.

The AB-Mountain produces free freshwater and gives very cheap energy.

**13. Using the AB-Dome for tourism, communication and wind energy.**

The relatively low AB-Dome can give good profit from tourism. As it shown in [9] p.93 for 4800 tourists in day and ticket cost \$9 the profit would be about \$15 million/year. (They would go for the sweeping horizon-to horizon view of a vast terrain kilometers below)

The profit may be from communication (TV, cell-telephone, military radars, etc). The best profit would be from high altitude wind electric station [14] (for wind speed $V = 13$ m/s at $H = 4$ km, wind rotor $R = 100$ m the power will be about 20 MW). In reality the wind speed at this altitude is more strong (up 35 m/s), stable, and rotor may to have radius up 150 m. That means the wind power may reach from 10 - 30 times more power. Since the generated power can go up as the cube of the wind speed, the installation would be well amortized.

**14. Cost of material.** The cost some material are presented in Table 2 (2005-2007). Some difference in the tensile stress and density are result the difference sources, models and trademarks.

**Table 4**. Average cost of material

| Material | Tensile stress, MPa | Density, g/cm3 | Cost USD \$/kg |
|----------|---------------------|----------------|----------------|
| **Fibers:** | | | |
| Glass | 3500 | 2.45 | 0.7 |
| Kevlar 49, 29 | 2800 | 1.47 | 4.5 |
| PBO Zylon AS | 5800 | 1.54 | 15 |
| PBO Zylon HM | 5800 | 1.56 | 15 |
| Boron | 3500 | 2.45 | 54 |
| SIC | 3395 | 3.2 | 75 |
| Saffil (5% $SiO_2$+$Al_2O_3$) | 1500 | 3.3 | 2.5 |
| **Matrices:** | | | |
| Polyester | 35 | 1,38 | 2 |
| Polyvinyl | 65 | 1.5 | 3 |
| Aluminum | 74-550 | 2.71 | 2 |
| Titanum | 238-1500 | 4.51 | 18 |
| Borosilicate glass | 90 | 2.23 | 0.5 |
| Plastic | 40-200 | 1.5-3 | 2 - 6 |
| **Materials:** | | | |
| Steel | 500 - 2500 | 7.9 | 0.7-1 |
| Concrete | - | 2.5 | 0.05 |
| Cement (2000) | - | 2.5 | 0.06-0.07 |

# Projects

## 1. AB-Mountain ridge of height 5 km and length 100 km

Let us estimate the parameters, cost and profit of a system of AB-Mountains. Ridge of the height $H = 5$ km and length $L = 100$ km having simple primary cover (without control of transparency). Take the over pressure 0.01 atm (1000 N/m$^2$), safety tensile stress of cover $\sigma = 100$ kg/mm$^2$ ($10^9$ N/m$^2$), specific weight of cover $\gamma = 1500$ kg/m$^3$.

**Main cost of AB-Dome**. From Fig. 20 we find the thickness of the cover at ground is $\delta = 5$ mm, at 5 km is 3 mm. If we use the few support cable (decrease $R$ in Eq. (9)[1]), the need cover thickness significantly decreases. We take the average $\delta = 3$ mm. The weight of $S = 1$ m$^2$ cover is $w = \sigma\gamma S = 4.5$ kg/m$^2$. The length of dome semi-cylinder $L_r = \pi R = 1.57 \times 10^4$ m, the weight of 1 m dome length is $W_1 = wL_r = 7 \times 10^4$ kg/m.

If cost of 1 kg cover is \$1, then the cost of 1 m of the dome cover is \$70,000/meter). We take the **total** cost of 1 m Dome length (1 m of ready-made Dome Installation) $c_d = \$100,000$/meter or \$100 million/km. In this case the total cost of $L = 100$ km AB-Dome Installation is $C_d = c_d L = \$10^{10} =$ **\$10 B** (where B is billion dollars). If maintenance is \$1000/m/year, the total maintenance is \$100M/year (where M means million). The cost of cover is the main component in the AB-Mountain. If support cables are used the thickness (and cost) of cover may be decreased by some times (up to 10, [1]). The film may be reinforced by strong cheap artificial transparence fiber and this textile material will have very high tensile stress. The life time of the cover is about 20 years. (\$1 billion a year must be set aside for replacing it).

**Profit from AB-Dome**.

a) *Profit from sale of electricity*. Let us take the average annual temperature outside of Dome $T_0 = 25$ C. From fig.17 we find the average power of electricity is $P = 7$ MW/m. The Dome length is 100km $= 10^5$m. The total power is $7 \times 10^6 \times 10^5 = 7 \times 10^{11}$ W = 0.7 TW. The total (include oil, gas, coal, etc,) World Power usage was about 14 TW in 2005 (all World hydro-station has power only 0.3 TW). The full year contains $t = 24 \times 365 = 8,760$ hours. Assume the average wind blows 4 hours in day, and probability a needed direction is 0.25. Then useful year time is $t_u = 4 \times 0.25 \times 365 = 365$ hours. Annual energy of 1 m Dome is $E_1 = t_u P = 2.55 \times 10^6$ kWh/year/m. (2.55 billion kWh yearly for Mountain ridge 100 km).

Retail cost of electric energy in New York is \$0.18 1/kWh. We take price $c_1 = \$0.1$ 1/kWh. The profit from electricity of energy is $C_1 = c_1 E_1 = \$0.255$ M/year/m. The length of Dome is $L = 10^5$ m. The total annual profit only from electricity is $C_e = C_1 L = $ **\$25.5B/year**. *That is ~ 2.5 times more than the cost of the AB-Mountain installation.*

(In reality, of course, increased supply usually decreases prices; but even at \$.01 a kilowatt hour, the equivalent of coal at \$82 a ton or gas at under \$3 per thousand cubic feet, the income of \$25 billion would pay back the costs of construction (of the dome, not the hydro works) in *only five months!* And at half even this price, or a half-cent a kilowatt-hour, the cleanness and controllability of hydroelectricity would seriously start to displace carbon fuels, in the sense that even coal and gas would have difficulty competing as simple thermal sources. So an interesting initiative for peace in the Mediterranean area would be the EU financing such an AB-Mountain, the EU taking the power and the desert areas to the south taking the water.)

The actual profile of the available water may be distributed around the year, depending on when prevailing winds cause a local 'rainy season'

b) *Profit from sale of water*. For temperature $T_0 = 25$ C the water flow from 1 m Dome is $q = 0.17$ m$^3$/s/m, 1m$^3$ = 1kL (fig. 15). Full year has $t = 24 \times 60 \times 60 \times 365 = 3.1536 \times 10^7$ s. Assume the average wind blows 4 hours in day, and probability a needed direction is 0.25. Then useful year time is $t_u = $ 4×0.25×365 = 365 hours = 1.3×10$^6$ s. The annual water from 1 m Dome is $M_1 = qt_u$ = 2.2×10$^5$ kL/year/m. (2.2 ×10$^5$ ×10$^5$ = 2.2 × 10$^{10}$ tonnes = 22 cubic kilometers of fresh water!) (For comparison, the Egypt is only allowed to remove from the Nile, about 55.5 cubic kilometers per year. (http://www.eoearth.org/article/Water_profile_of_Egypt). If the Egyptian government were to build the AB-Mountain rather than to buy from a foreign water vendor, for \$10 billion a year and \$1 billion a year maintenance it could produce and sell the water itself to farmers and use the revenue to pay for government employees in the cities. The annual water from 1 m Dome is $M_1 = qt$ = 2.2×10$^5$ kL/year/m. If water cost is $c$ = \$1/kL and length of Dome $L = 10^5$ m, the total water profit from Dome will be **$C_w$ = \$22B/year**. *That is in 2.2 times more than the cost of the installation.*

  *The common annual income from electricity and water is* **\$47.5B/year**. *That is about 4.7 times more than the cost of the AB-Mountain installation.*

  Again, at one-twentieth or less these prices, (the likely price stagnation point) the AB-Mountain abundantly repays its construction, and in fact cultivates huge new demand for carbon replacement power and water cheap enough to irrigate with.

  c) *The profit from cooling of buildings*. In point 8 (Eq.(8)) we computed the example which shows that cooling water has $Q_1 = 38$MW/m for air temperature $T = 35^{\circ}$C. That may saves $Q_2 = 84$ MW/m electric energy spent by conditions. The year contains $t = 24 \times 365 = 8,760$ hours (working time $t_u = 365$ hours). Annual energy of 1 m Dome is $E_1 = t_uQ_1 = 1.39 \times 10^7$ kWh/year/m. If we sale this energy by price $c_1 =$ \$0.05/kWh the profit is $C_l = c_1E_1 =$ \$0.7M/year/m. The length of Dome is $L = 10^5$ m. The total annual income is $C_e = C_lL$ would  theoretically= \$70 B/year. *That is a benefit 7 times more than  the cost of installation.* That income is problematical; therefore we do not take it into account. (The offered method of cooling is new and with great benefit but the poor tropical countries that need it most cannot yet pay for it. However, as a side benefit (i.e. cooling poor tropical countries such as India where extremes of 48 C (120 F) are not uncommon for weeks on end in summer, it could save many lives. It could also relieve inestimable human suffering and buy political peace with the poorest of the poor by giving them a real benefit from their country's decision to build an AB-Mountain range. There are no technical problems in its application, the current water heat system may be used for cooling, and building cooling is very important problem for hot country or warm countries in hot summer weather!)

d) *Profit from harvest*. The surface into Dome will has excellent control climate and may be used for harvest [1] three times in year (hydroponics). The area under cover is $S = 10^9$ m$^2$. If profit is $c =$ \$1/m$^2$, the common annual profit is $C_y = cS =$ **\$3B/year**.

e) *Profit from rent*. The area under Dome cover is beautiful place for city, because has a fine warm weather all time, may be protected from nuclear warhead, chemical, and biological weapon in war time [16]. If additional rent for ground into Dome is $c =$ \$1/month/m$^2$, the annual profit will be $C_r = 12cS = 12 \times 1 \times 10^9 =$ **\$12B/year**.

f) *There are a lot of other possibilities* to get the profit from offered AB-Mountain. For example, tourism at high altitude, communication, civil or/and military locators, fly gliding, cheap parachute jumping, non-gravity jumping, entertainments, etc. That can give additional profit.

  As you see the AB-Mountain may just qualify as the project in the World that best gives incomparable rates of return, earning its cost back and more in one year (every year!). Whoever does not agree, can do the numbers with their own assumptions, to check our own.

  Naturally, any national or private investor would have many questions: For example, how to build this gigantic Dome. The author has many inventions, which solve these problems and is available for consulting to interested parties.

  We took the height 5 km. However, the artificial AB-Mountain having the height 3.5 – 4 km will have a closed efficiency, but lesser cost in 30 – 40%.

## DISCUSSION

As with any innovative macro-project proposal, the reader will naturally have many questions. Author offer brief answers to the four most obvious questions our readers are likely to ponder.

(1) *How can snow and ice be removed from the dome?*

If water appears over film (rain), it flows down through a special tube into ground-based turbines. If snow (ice) appears atop the film, the control system passes the warm air between two cover layers. The warm air melts the snow (ice) and water flows down. The film cover is flexible and has a lift force of about 20 - 200 kg/m$^2$.

(2) *Storm wind.*

The Dome has special semi-cylindrical form (fig.8, 9). For internal pressure is 0.01 atm the Dome can resist a storm wind up to 40 m/s (144 km/hour). If wind is more powerful yet, we can increase the internal pressure up to the needed value.

(3) *Cover damage.*

The envelope contains a rip-stop cable mesh so that the film cannot be damaged greatly. Electronic signals alert supervising personnel of any rupture problems. The cover has internal and external cable (rope-ladders and rope cars, rope elevators) and workers can reach the any part of cover inside or out side for repair.

(4) *What is the design life of the film covering?*

Depending on the kind of materials used, it may be as much a decade (or up to 20 years and more). In all or in part, the cover can be replaced periodically.

(5) *How to build the AB-Dome?* The simplest method is to spread the section of cover on ground and turn on the pump. Author also has developed many specialized submethods.

The author began this research as investigation of new method for receiving the cheap freshwater from atmosphere. In processing research, he discovered that method allows producing huge amount energy, in particular, by transferring the atmospheric energy into electricity with high efficiency. The thin film (relative to volume contained, and absolutely) is very cheap. They are thrown out by the hundreds of tons every day and waste the environment, but properly employed they can conserve it as well. The theory of inflatable space towers [1]-[16] allows to build very cheap high height AB-Domes or towers, which can be used also for tourism, communication, radio-location, producing wind electricity, space research [1-16].

## 5. CONCLUSION

One half of Earth's population is malnourished. *The majority of Earth is not suitable for unshielded human life.* The offered AB-Mountains can change the climate many regions, give them the water and energy. The increasing of agriculture area, crop capacity, carrying capacity by means of converting the deserts, desolate wilderness, taiga, permafrost into gardens are an important escape hatch from some of humanity's most pressing problems. The offered cheapest AB method may dramatically increase the potentially realizable sown area, crop capacity; indeed the range of territory suitable for human living. In theory, converting all Earth land such as Alaska, North Canada, Siberia, or the Sahara or Gobi deserts into prosperous garden would be the equivalent of colonizing an entire new planet. The suggested method is very cheap and may be utilized at the present time. We can start from small areas, such as small towns in bad regions and extended the practice over a large area—and what is as important, making money most of the way.

Film domes can foster the fuller economic development of dry, hot, and cold regions such as the Earth's Arctic and Antarctic and thus, increase the effective area of territory dominated by humans. The country can create the Mountain barriers which will defense country from cold North winds. Normal human health can be maintained by ingestion of locally grown fresh vegetables and healthful "outdoor" exercise. The domes can also be used in the Tropics and Temperate Zone. Eventually, they may find application on the Moon or Mars since a vertical variant, inflatable towers to outer space, are soon to become available for launching spacecraft inexpensively into Earth-orbit or interplanetary flights [12].

The related problems are researched in references [1]-[16].

Let us shortly summarize some advantages of this offered AB Dome method of climate moderation:

(1) The artificial Mountains give a lot of freshwater and energy and change a local climate (convert the dry climate to damp climate);

(2) They protect from cool or hot wind the large region;

(3) Covered area does not need large amounts of constant input water for irrigation;

(4) Low cost of inflatable film Dome per area reclaimed;

(5) Control of inside temperature;

(6) Usable in very hot and cool regions;

(7) Covered area is not at risk from weather;

(8) Possibility of flourishing crops even with a sterile soil (hydroponics);

(9) 2 – 3 harvests in year; without farmers' extreme normal risks.

(10) Rich harvests, at that.

(11) Converting deserts, desolate wilderness, taiga, tundra, permafrost, and ocean into gardens;

(12) Covering the towns, cities by offered domes;

(13) Protection of city from external, tactical nuclear warhead, chemical and biological weapon [16];

(14) Using the high artificial Mountains for tourism, communication, long location, and so on;

(15) Using the dome cover for illumination, pictures, films and advertising.

We can make fine local weather, get new territory for living with an agreeable climate without daily rain, wind and low temperatures, and for agriculture. We can cover by thin film gigantic expanses of bad dry and cold regions. The countries having big territory (but bad land) may be able to use to increase their population and became powerful states in the centuries to come.

The offered method may be used to conserve a vanishing sea as the Aral or Dead Sea. A closed loop water cycle saves this sea for a future generation, instead of bequeathing a salty dustbowl.

The author developed the same method for the ocean (sea). By controlling the dynamics and climate there, ocean colonies may increase the useful area another 3 times (after the doubling of useful land outlined above) All in all, this method would allow increasing the Earth's population by 5 – 10 times without the starvation.

The offered method can solve the problem of **global warming** because AB domes will be able to confine until use much carbonic acid ($CO_2$) gas, which appreciably increases a harvest. This carbon will show up in yet more productive crops! The dome lift force reaches up 300 kg/m$^2$. The telephone, TV, electric, water and other communications can be suspended to the dome cover.

The offered method can also help to defend the cities (or an entire given region) from rockets, nuclear warheads, and military aviation. Details may be offered in a paper [16].

### Acknowledgement

The author wishes to acknowledge Joseph Friedlander for correcting the author's English and useful advice and suggestions.

Version 1 is submitted on 31 Jan 2008, small corrected version 2 is submitted on 10 May 2008.